\documentclass[journal=pc,manuscript=article]{achemso}
\usepackage{longtable}
\usepackage{multirow}
\usepackage{amsmath}
\usepackage{subfigure}
\usepackage[usenames,dvipsnames]{xcolor}
\usepackage{soul}
\usepackage{bm}

\usepackage{amssymb}
\usepackage{multicol} \usepackage{wrapfig} \usepackage{enumitem}
\usepackage{bm} \usepackage{outlines} %for itemize with subitems
\usepackage[normalem]{ulem}

\usepackage{microtype}

\usepackage{url}
\usepackage{hyperref}

\usepackage{xr}

\makeatletter
\newcommand*{\addFileDependency}[1]{% argument=file name and extension
  \typeout{(#1)}
  \@addtofilelist{#1}
  \IfFileExists{#1}{}{\typeout{No file #1.}}
}
\makeatother

\newcommand*{\myexternaldocument}[1]{%
    \externaldocument{#1}%
    \addFileDependency{#1.tex}%
    \addFileDependency{#1.aux}%
}
\myexternaldocument{si}

\SectionNumbersOn

\author{Sena Aydin} \affiliation[University of Basel]{Department of
  Chemistry, University of Basel, Klingelbergstrasse 80, CH-4056
  Basel, Switzerland.}\altaffiliation{These authors contributed
  equally}

\author{Seyedeh Maryam Salehi} \affiliation[University of
  Basel]{Department of Chemistry, University of Basel,
  Klingelbergstrasse 80, CH-4056 Basel, Switzerland.}
\altaffiliation{These authors contributed equally}

\author{Kai T\"{o}pfer} \affiliation[University of Basel]{Department
  of Chemistry, University of Basel, Klingelbergstrasse 80, CH-4056
  Basel, Switzerland.}

\author{Markus Meuwly} \affiliation[University of Basel]{Department of
  Chemistry, University of Basel, Klingelbergstrasse 80, CH-4056
  Basel, Switzerland.} \alsoaffiliation{Department of Chemistry, Brown
  University, RI 02912, USA} \email{m.meuwly@unibas.ch}

\title {SCN as a Local Probe of Protein Structural Dynamics}

\begin{document}
\date{\today}

\begin{abstract}
The dynamics of lysozyme is probed by attaching -SCN to all
alanine-residues. The 1-dimensional infrared spectra exhibit frequency
shifts in the position of the maximum absorption by 4 cm$^{-1}$ which
is consistent with experiments in different solvents and indicates
moderately strong interactions of the vibrational probe with its
environment. Isotopic substitution 
$^{12}$C $\rightarrow ^{13}$C leads to a red-shift by 
$-47$ cm$^{-1}$ which is consistent with
experiments with results on CN-substituted copper complexes in
solution. The low-frequency, far-infrared part of the protein spectra
contain label-specific information in the difference spectra when
compared with the wild type protein. Depending on the positioning of
the labels, local structural changes are observed. For example,
introducing the -SCN label at Ala129 leads to breaking of the
$\alpha-$helical structure with concomitant change in the far-infrared
spectrum. Finally, changes in the local hydration of SCN-labelled
Alanine residues as a function of time can be related to angular
reorientation of the label. It is concluded that -SCN is potentially
useful for probing protein dynamics, both in the high-frequency
(CN-stretch) and far-infrared part of the spectrum.
\end{abstract}

\section{Introduction}
Characterizing the dynamics of proteins in solution is essential for
understanding their function and interactions with their environment
which is particularly relevant in a physiological and cellular
context.\cite{schuler:2017} Because functional motions of proteins can
involve local and/or more global changes in their structure and single
protein structures are not sufficient to explain a protein's
function,\cite{lane:2023} it is or particular interest to delineate
protein structural changes on given time scales. One technique to link
structure, dynamics and the time scales on which they develop is
optical spectroscopy, in particular infrared (IR)
spectroscopy.\cite{yang:2015,fayer:2012,gai:2001,raleigh:2010,MM.rev.jcp:2020}
Recent progress in synchronization of femtosecond laser pulses allowed
to map out the entire reaction cycle of
bacteriorhodopsin.\cite{hamm:2023}\\

\noindent
To gain site-specific information on local protein dynamics,
significant effort has been focused on developing chemical groups
suitable for frequency-resolved
spectroscopy.\cite{schultz:2006,gai:2015,cho:2013} For instance, -CD
bonds have been shown to be suitable to map out the local dynamics
within a protein.\cite{intro:2009} In another work, attaching a
ruthenium carbonyl complex to the His15 residue in both hen egg white
and human lysozyme, allowed to investigate the local hydration of the
proteins via IR spectroscopy.\cite{kubarych:2012} Finally, $^{13}$C as
a site-specific isotopic label was introduced to examine the secondary
structures of the residues in conformationally heterogeneous
peptides.\cite{walters:1991} Using computer simulations it has been
shown that AlaN$_3$ is a position-sensitive probe, providing useful
information on the protein modification site in previous
work.\cite{MM.lys:2021} \\

\noindent
Cyanide (-CN) and thiocyanate (-SCN) are nitrile-based vibrational
probes that can be integrated into proteins and have been used, for
example, to label alanine
residues.\cite{edelstein:2010,londergan.jpcl:2010,londergan.bj:2010}
When the -SCN moiety is incorporated into peptides or proteins, it
gives rise to absorption bands owing to its comparatively large
extinction coefficient between 100 to 300
M$^{-1}$cm$^{-1}$.\cite{boxer:2006,gai:2011} The IR stretching
absorption values of SCN-labelled proteins change between 2151 and
2161 cm$^{-1}$ and it is sensitive to its environment acting as a
site-specific electric field probe for proteins.\cite{boxer:2006} The
-SCN probe has also been attached to alanine to form
$\beta-$thiocyanatoalanine for which
NMR\cite{malthouse:1996,londergan:2007} and IR studies in solution had
been carried out. Moreover, the lifetime of the C--N stretching mode
in a -SCN label is sensitive to the surrounding environment, owing to
the decoupling of the CN-label and the rest of the amino acid due to
the sulfur atom, which has been referred to as the ``insulating
effect''.\cite{cho:2013,bredenbeck:2020} Similar results were observed
in a study of methyl thiocyanate (MeSCN) in different
solvents.\cite{bredenbeck:2014}\\

\noindent
In the present work, -SCN is used as a spectroscopic probe to
elucidate local protein dynamics. The label is attached to all 16
alanine residues of the lysozyme protein to investigate the
differences in the local and global dynamics of the protein.  First,
the methods are described, followed by a validation of the potential
energy surfaces (PES). Next, the protein IR spectroscopy and the
dynamics of the spectroscopic probe is described for lysozyme with all
alanine-residues labelled with -SCN.  Finally, the results are
discussed and conclusions are drawn.\\

\section{Methods}

\subsection{The Potential Energy Surface} 
\begin{figure}[H]
\begin{center}
\includegraphics[width=\textwidth]{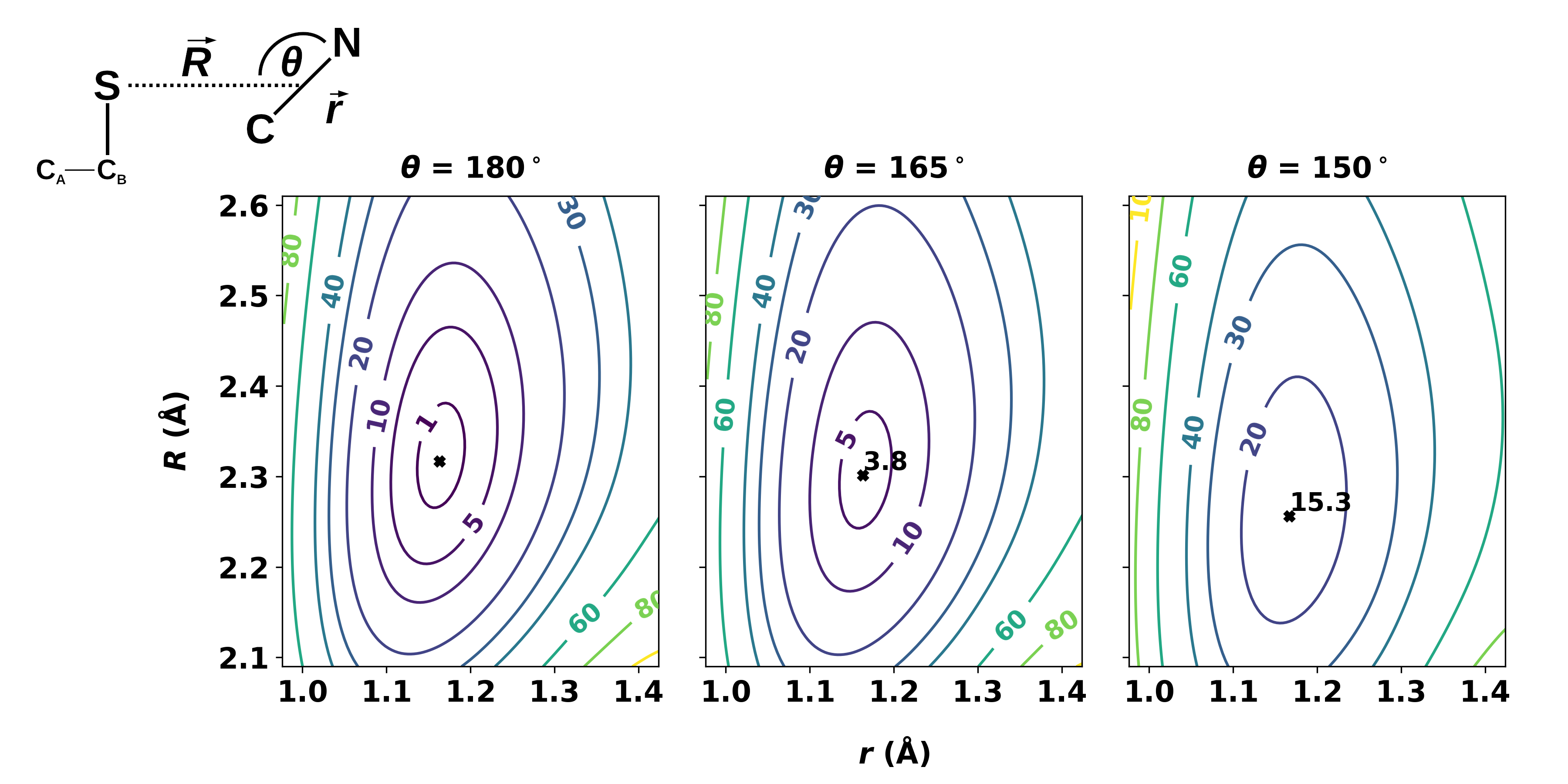}
\caption{Contour representation of the RKHS PES for -SCN label based on
  PNO-LCCSD(T)-F12 {\it ab initio} points in Jacobi coordinates $(R,
  r, \theta)$ for $\theta=180.0^\circ$, $\theta=165.0^\circ$, and
  $\theta=150.0^\circ$, from left to right. The black cross in the center of each panel
  shows the minima of the PES at the fixed angle in kcal/mol (for
  $\theta=180.0^\circ$ it is 0 kcal/mol). The schematic representation
  of -SCN Jacobi coordinates are given in the left corner of the figure.}
\label{fig:pes}
\end{center}
\end{figure}

The energy function is integrated with CHARMM\cite{charmm:1998} force
field and the regenerating kernel Hilbert space
(RKHS)\cite{RKHS-Rabitz-1996,MM.rkhs:2017} PES to probe spectroscopic
methods for the protein structure.  Starting from an optimized
structure of thiocyanatoalanine at the MP2/aug-cc-pVTZ level of
theory, the PES for the -SCN label is calculated using pair natural
orbital-based coupled cluster level
(PNO-LCCSD(T)-F12)\cite{lccsd-schwilk-2017,lccsdf12-schwilk-2017}
together with the aug-cc-pVTZ basis set\cite{DUN:JCP89} using MOLPRO
software.\cite{molpro2} The \textit{ab initio} energies were
calculated in Jacobi coordinates $(R, r, \theta)$, where $r$ is the
distance between carbon and nitrogen atoms, $R$ is the distance
between their center of mass and sulfur atom, and $\theta$ is the
angle between $\vec{r}$ and $\vec{R}$, see the inset of Figure
\ref{fig:pes}. The angular grid ($\theta$) used here contains 6 points
between 150 to $180^\circ$ while radial grids include 16 points along
$r$ ranging from 0.92 to 1.68 \AA\/ and 13 points along $R$ between
1.82 and 2.82 \AA. A dihedral angle $\phi$ is defined between the
planes spanned by $\vec{r}$ and $\vec{R}$ of the -SCN label and the
vector from the anchor carbon atom of alanine and the sulfur atom of
the -SCN label with $\vec{R}$. That dihedral angle $\phi$ is preserved
for the energy calculations as in the equilibrium conformation
$\phi_\mathrm{eq} = 11.3^\circ$.  The equilibrium Jacobi angle
$\theta$ is $\sim 180^\circ$.\\

\noindent
For the differentiable representation of the {\it ab initio} energies
using the RKHS approach\cite{RKHS-Rabitz-1996,MM.rkhs:2017} only
coupled cluster results with a T1 diagnostic value\cite{T1diagnostic}
lower than $0.02$ are used. The condition excludes, among others, all
reference points with $r > 1.4$ \AA\/ for the construction of the RKHS
potential. Still, the included reference points cover the conformation
space with respective potential energies up to 70 kcal/mol around the
minimum conformation. Figure \ref{fig:pes} demonstrates the RKHS
representation of the PES constructed from 842 of the 1248 {\it ab
  initio} energies. Note that the PES at $r > 1.4$ \AA\/ is physically
not meaningful due to the exclusion conditions.\\

\subsection{Molecular Dynamics Simulations}
Molecular Dynamics (MD) simulations for the wild type (WT) and all
modified AlaQQSCN with QQ = 41, 42, 49, 63, 73, 74, 82, 93, 97, 98,
112, 129, 130, 134, 146, and 160 were performed using the
CHARMM\cite{charmmFF22} package starting from the X-ray structure
1L83.\cite{morton:1995}.  To carry out the simulation with the RKHS
PES for the spectroscopic label, an interface is
written.\cite{MS.n3:2019} For the -SCN label a point charge
representation based on a natural bond orbital (NBO) analysis of the
minimized MP2/aug-cc-pVTZ structure of thiocyanatoalanine was used. The total
charge of the -SCN label from this analysis was close to 0 with
rescaled partial charges $q_{\rm S} = 0.11e$, $q_{\rm C} = 0.02e$,
$q_{\rm N} = -0.13e$.  The scaling was determined from comparing the
NBO charges for the alanine residue with those from the CGenFF
parametrization.  The geometry of the label was defined by a bond
between the sulfur atom and the side chain methyl-carbon atom C$_{\rm
  B}$ of alanine, the corresponding CSC$_{\rm B}$ angle, and the
SC$_{\rm B}$C$_{\rm A}$ angle but without including the NCSC$_{\rm B}$
dihedral angle, see below.\\

\begin{figure}[H]
\begin{center}
\includegraphics[width=0.9\textwidth]{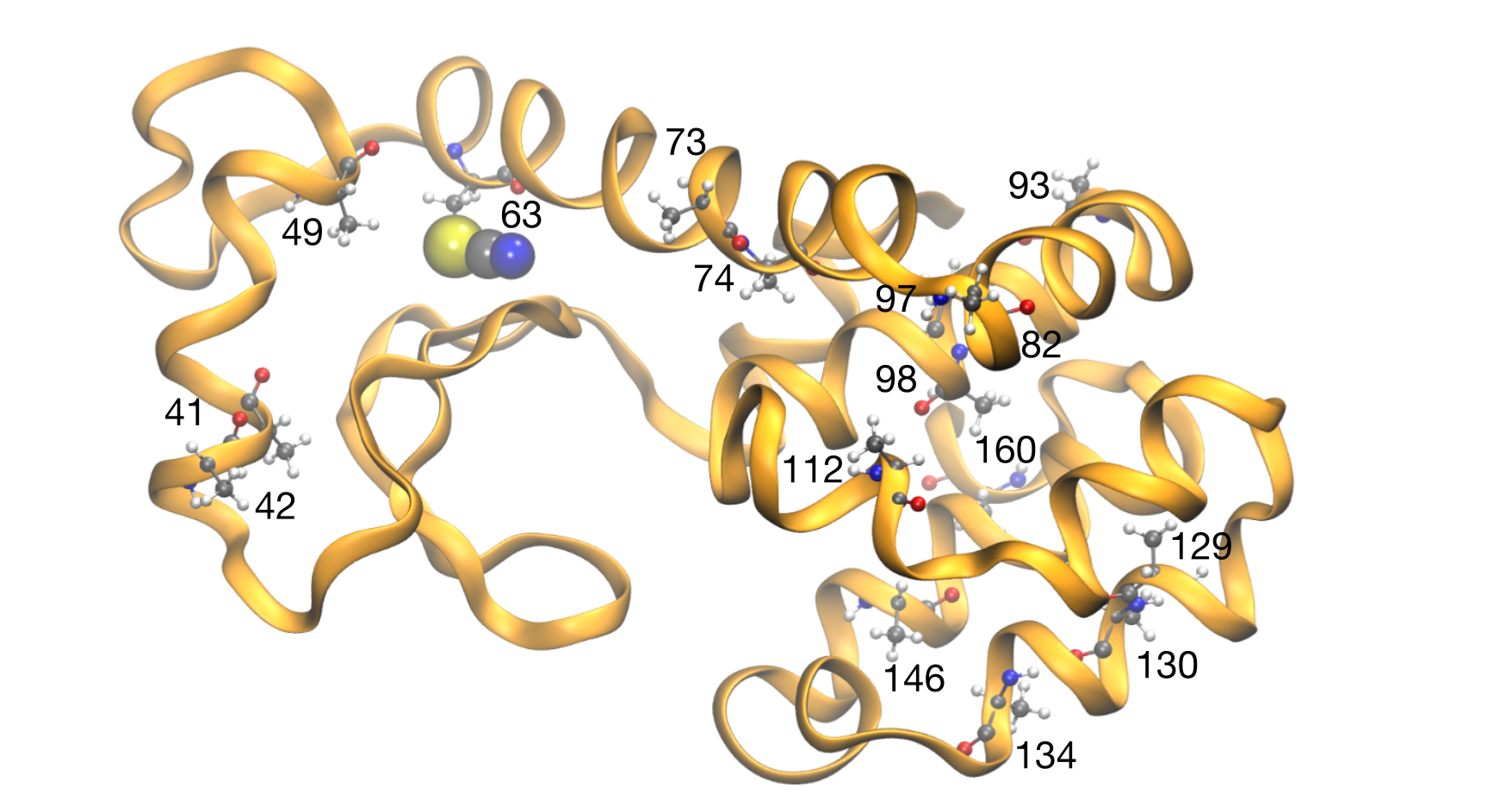}
\caption{Structure of lysozyme emphasizing the positions of 16
  labelled alanine residues. The labelled alanine residues are located
  at 41, 42, 49, 63, 73, 74, 82, 93, 97, 98, 112, 129, 130, 134, 146,
  and 160. SCN-labelled Ala63 is shown as an example.}
\label{fig:structure}
\end{center}
\end{figure}

\noindent
The simulations were carried out in a cubic box of size $(78)^3$
\AA\/$^3$ with explicit TIP3P water model. Using the
SHAKE\cite{SHAKE-Gunsteren-1997} algorithm, bond lengths involving
H-atoms were constrained. A cutoff of 14 \AA\/ with switching at 10
\AA\/ was used for non-bonded interactions.\cite{Steinbach1994} In
Figure \ref{fig:structure}, the structure of lysozyme with 16 labelled
alanine residues is illustrated. The -SCN label is attached to each of
the 16 alanine residues. For all 16 AlaQQSCN modifications, 2 ns
production run was performed in the $NVT$ ensemble after the initial
minimization, heating, and equilibration.\\

\subsection{Spectroscopic Analysis}
For the protein and subsystems 1-dimensional IR spectra $I(\omega)$
were obtained from the Fourier transform of the dipole-dipole
correlation function
\begin{equation}
  I(\omega) n(\omega) \propto Q(\omega) \cdot \mathrm{Im}\int_0^\infty
  dt\, e^{i\omega t} \sum_{i=x,y,z} \left \langle
  \boldsymbol{\mu}_{i}(t) \cdot {\boldsymbol{\mu}_{i}}(0) \right
  \rangle
\label{eq:IR}
\end{equation}
where $\boldsymbol{\mu}_{i}(t)$ is the dipole moment vector of the
full protein, the modified alanine residue, or the -SCN label. A
quantum correction factor $Q(\omega) = \tanh{(\beta \hbar \omega /
  2)}$ was applied to the results of the Fourier
transform.\cite{marx:2004} This procedure yields lineshapes but not
absolute intensities.\\

\noindent
As a proxy of the 2-dimensional spectroscopy the frequency fluctuation
correlation function (FFCF) was determined.  For this, the frequency
trajectory $\omega(t)$ follows the temporal change of the
instantaneous CN-normal mode of all 16 AlaQQSCN was calculated for
$4\times10^5$ snapshots of each production simulation using
instantaneous normal mode (INM) analysis.\cite{MS.insulin:2020} For
each snapshot, the structure of the -SCN label was minimized with
frozen environment. From the frequency trajectory $\omega(t)$ the FFCF
of $\delta \omega(t) = \omega(t) - < \omega >$ was determined which
contains valuable information on relaxation time scales corresponding
to the solvent dynamics around the solute. The FFCFs are fit to an
empirical expression
\begin{equation}
  \langle \delta \omega(t) \delta \omega(0) \rangle = a_{1}
  \cos(\gamma t) e^{-t/\tau_{1}} + a_{2}
  e^{-t/\tau_{2}} + \Delta_0^2
\label{eq:ffcffit}
\end{equation}
which allows analytical integration to obtain the lineshape
function\cite{hynes:2004} using an automated curve fitting tool from
the SciPy library.\cite{2020SciPy-NMeth} Here, $a_{i}$, $\tau_{i}$,
$\gamma$ and $\Delta_0^2$ are the amplitudes, decay time scales, phase
and asymptotic value of the FFCF. The first time is included in order
to model early-time undulations in the FFCF which have been
interpreted as solvent-solute
interactions.\cite{skinner:2006,hynes:2004,MM.cn:2013,MM.nma:2014}\\

\subsection{Local Hydrophobicity}
Calculating the local hydrophobicity (LH) which is defined as a time
dependent quantity, $\delta\lambda_{\textrm{phob}}^{(r)}(t)$, is an
efficient way to quantify the solvent exposure of amino
acids.\cite{Willard-jctc-2018,Willard-jpcb-2018} The method analyzes
the occupation and orientational statistics of surface water molecules
at the protein/water interface using the vector $\vec{\kappa} = (a,
\cos \theta_{\rm OH1}, \cos \theta_{\rm OH2})$. Here, $a$ is the
distance between the nearest atom of residues with oxygen of water
molecules, and $\theta_{\rm OH1}$ and $\theta_{\rm OH2}$ are the
angles between the water OH1 and OH2 bonds and the interface
normal.\cite{MM.hb:2020} Defining the local hydrophobicity (LH) as
$\delta \lambda_\mathrm{phob}^{(r)}(t) =
\lambda_\mathrm{phob}^{(r)}(t) - \langle \lambda_\mathrm{phob}
\rangle_0$, then
\begin{equation}
    \lambda_\textrm{phob}^{(r)}(t) = -\frac{1}{{\sum_{a=1}^{N_a(r)}
        N_w(t;a)}}\sum_{a=1}^{N_a(r)} \sum_{i=1}^{N_w(t;a)} \ln{\left[
        \frac{P(\vec{\kappa}^{(i)}(t)|\textrm{phob})}
             {P(\vec{\kappa}^{(i)}(t)|\textrm{bulk})} \right]}
\label{eq:lh}
\end{equation}
and $\langle \lambda_\mathrm{phob} \rangle_0$ is the ensemble average
for the ideal hydrophobic reference system. $N_a(r)$ refers to all
atoms in residue $r$ while $N_w(t;a)$ stands for all water molecules
within a cut-off of 6 \AA\/ of atom $a$ at time $t$.\cite{MM.hb:2020}
The vector $\vec{\kappa}^{(i)}(t)$ defines the $i$th water molecule in
the sampled population.\\

\noindent
The $P(\vec{\kappa}^{(i)}(t)|\textrm{phob})$ distribution is based on
hydrophobic reference system while
$P(\vec{\kappa}^{(i)}(t)|\textrm{bulk})$ is obtained from the actual
simulation.\cite{Willard-jctc-2018} From this perspective the local
hydrophobicity determines the similarity of the water ordering between
an ideal hydrophobic system and the one from the simulation. The more
dissimilar the two distributions are, the more hydrophilic the
environment of the site is and vice versa. Therefore, $\delta
\lambda_\mathrm{phob}^{(r)}(t) \approx 0$ indicates a hydrophobic
environment around residue $r$ while
$\delta\lambda_\mathrm{phob}^{(r)} > 0.5$ points to a more hydrophilic
environment.  However, the cutoff magnitude may change depending on
the system.\cite{Willard-jctc-2018,Willard-jpcb-2018,MM.hb:2020}\\

\section{Results}

\subsection{Validation of the PES}
\begin{figure}[H]
    \centering
    \includegraphics[width=16cm]{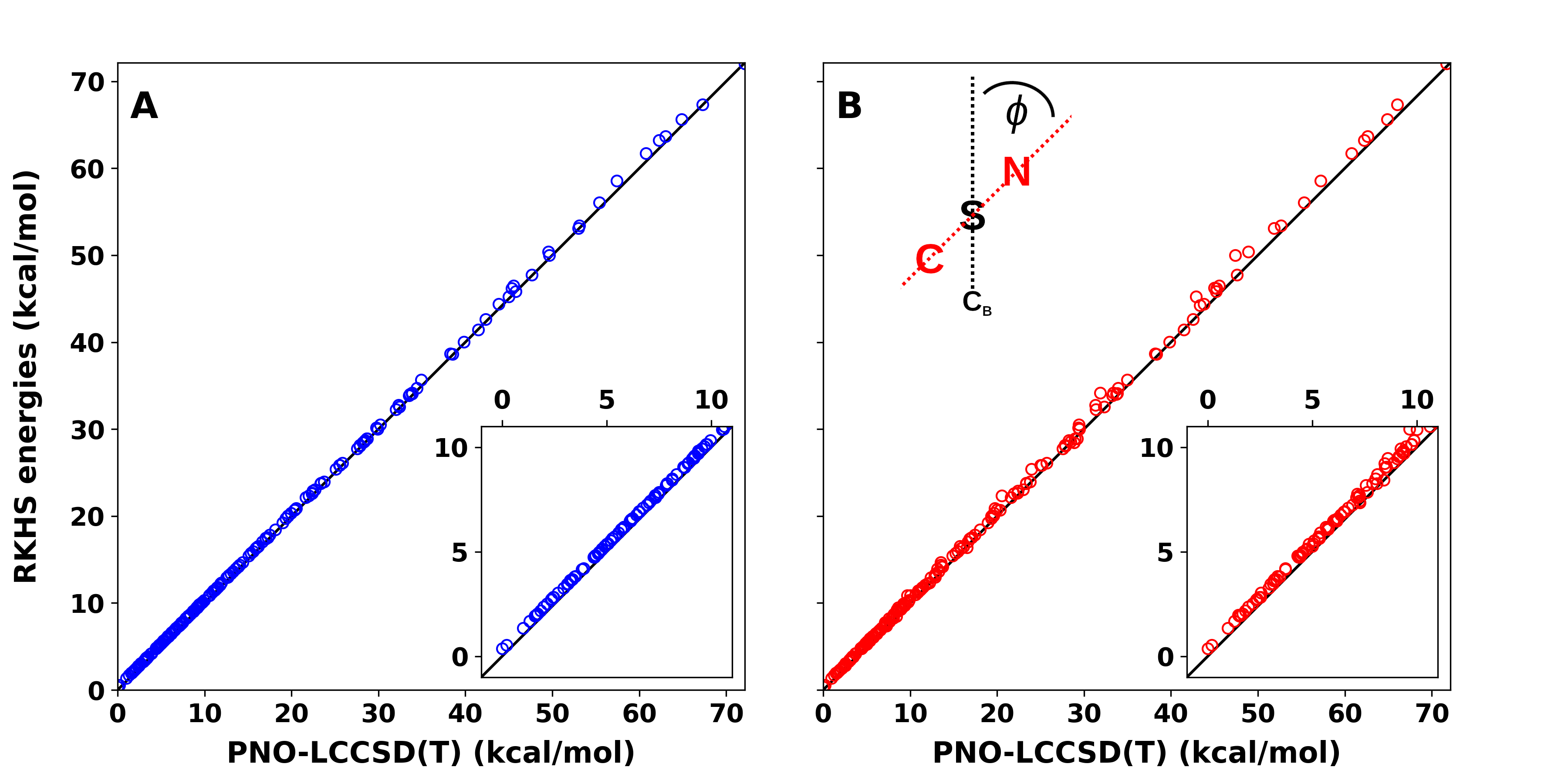}
    \caption{Correlation between PNO-LCCSD(T)/aug-cc-pVTZ reference
      {\it ab initio} energies and the RKHS-representation. Panel A:
      On-grid data for the dihedral fixed at the minimum energy
      structure ($\phi = 11.3^\circ$) and panel B: for a randomly
      oriented label with $\phi \in [0, 360]^\circ$). For panel A:
      $R^2 = 0.999$ and RMSD = 0.384 kcal/mol and for panel B: $R^2 =
      0.998$ and RMSD = 0.652 kcal/mol.  }
    \label{fig:offgrid}
\end{figure}

The 3-dimensional RKHS representation $V(R,r,\theta)$ of the -SCN label
as a function of the Jacobi coordinates $\{R, r, \theta \}$ reproduces
the reference calculations at the PNO-LCCSD(T)-F12/aug-cc-pVTZ level
of theory with an RMSD of $\sim 0.4$ kcal/mol ($R^2 = 0.999$), see
Figure \ref{fig:offgrid}A.  The geometry was that of the minimum
energy structure for thiocyanatoalanine for which all internal
coordinates were frozen except for the Jacobi coordinates describing
the geometry of the -SCN label. The energies cover a range within 70
kcal/mol of the global minimum.  To further validate the quality of
the PES, an additional 200 {\it ab initio} energies were determined
for off-grid geometries for given C$_{\rm A}$SCN dihedral angle
$\phi$, see Figure \ref{fig:offgrid} for the definition of $\phi$.
Geometries with \{$R, r, \theta$\} are randomly chosen by normal
distribution around \{$2.323$\,\AA, $1.178$\,\AA, $180^\circ$\} with a
standard deviation of \{$0.15$\,\AA, $0.1$\,\AA, $10.0^\circ$\} and
the dihedral angle was uniformly sampled in the interval $[0,
  360]^\circ$.  Figure \ref{fig:pes}B confirms the quality of the
RKHS-PES with ${\rm RMSD} = 0.652$ kcal/mol and $R^{2} = 0.998$.\\

\subsection{Protein and Sub-System Infrared Spectroscopy}
\begin{figure}[H]
\begin{center}
\includegraphics[width=0.9\textwidth]{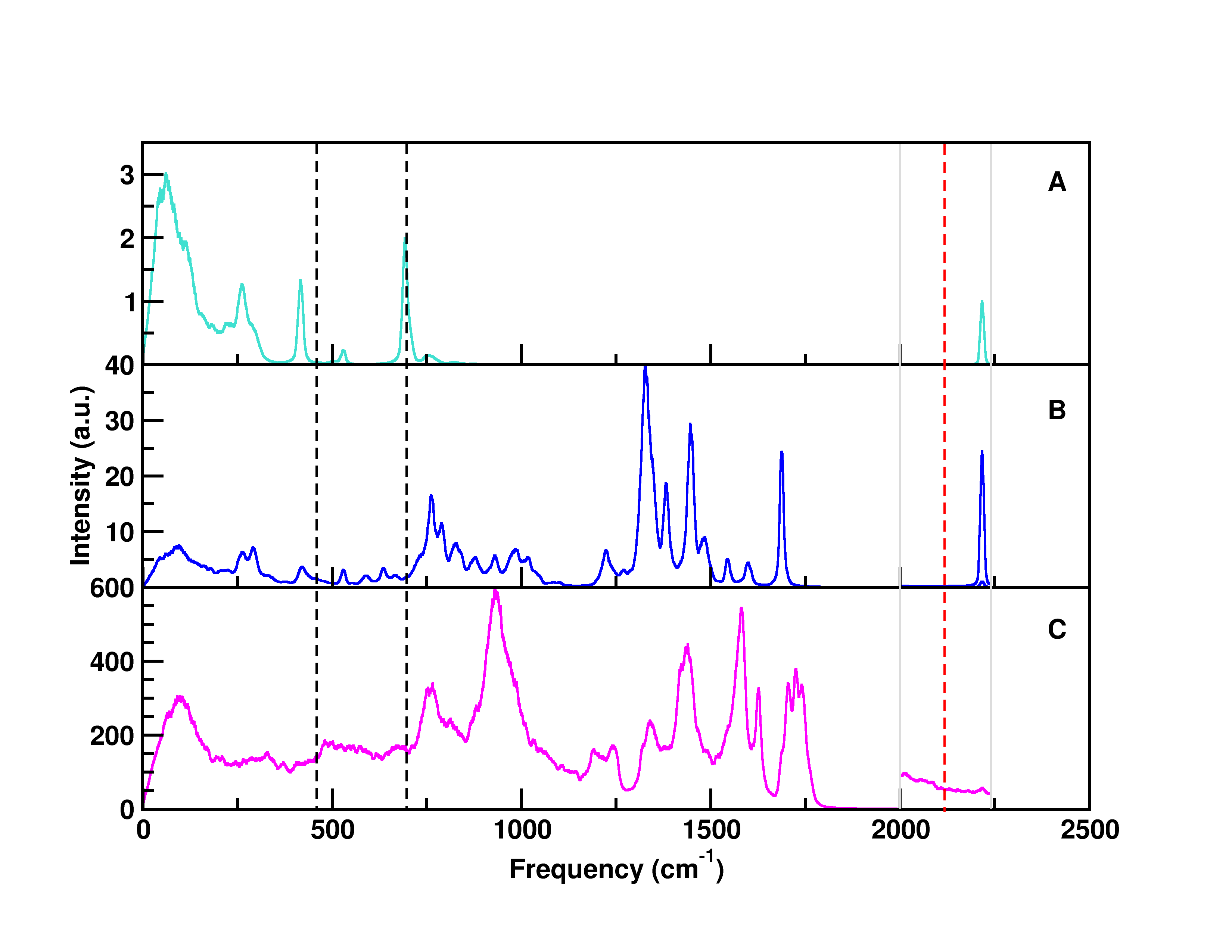}
\caption{IR spectra of -SCN label (A), labelled Ala93 (B), and full
  lysozyme (C) for Ala93SCN in lysozyme. The dashed vertical lines denote the
  specific peaks of -SCN label MeSCN at 459, 697, and 2118
  cm$^{-1}$. The range highlighted with
  grey lines between 2000 and 2240 cm$^{-1}$ is rescaled. The
  fundamental frequencies of Ala93SCN are at 414, 690, and 2217
  cm$^{-1}$.}
\label{fig:ALA93}
\end{center}
\end{figure}

First, the 1-dimensional IR spectra of the protein, the modified
Ala-residues, and the label were analyzed from the Fourier transform
of the dipole-dipole correlation function, see Figure \ref{fig:ALA93} for
labelled Ala93. The label-specific, Ala93SCN, and full protein
spectra are reported in panels A to C and the vertical dashed lines
indicate the bend, symmetric stretch, and asymmetric stretch vibration
of -SCN as assigned from the figure.\\

\noindent
Figure \ref{fig:ALA93}A shows the label-specific IR spectrum
determined from the sub-system dipole moment of the -SCN label. Besides
the clearly visibly internal vibrational modes (bend, symmetric
stretch, and asymmetric stretch) additional bands in particular
towards lower frequencies arise. Shifted to lower frequencies,
additional spectral features arise which represent coupling between
the motion of the label and the protein and solvent environment. For
Ala93SCN, the peaks are at 2217, 414, and 690 cm$^{-1}$.\\

\noindent
The IR spectrum of the Ala93SCN subsystem is reported in Figure
\ref{fig:ALA93}B. The SCN-asymmetric stretch vibration is well removed
from the amide, fingerprint and low-frequency vibrations and clearly
visible whereas the symmetric stretch disappears entirely. On the
other hand, the bending vibration is still visible. This partial IR
spectrum also clarifies that the low-frequency modes to which -SCN and
Ala93 couple are comparable. In the frequency range above 1000
cm$^{-1}$ a series of distinct vibrational peaks arises which could be
assigned if needed.\\

\noindent
Entire proteins have a yet larger density of states and discerning
particular features in the vibrational spectroscopy below 2000
cm$^{-1}$ becomes very challenging, see Figure \ref{fig:ALA93}C. A
significant increase in the number of vibrational modes is found and
only the -SCN asymmetric stretch can still be distinguished. Due to
the coupling between the various degrees of freedom many peaks wash
out and broaden considerably. However, the overlap for the
low-frequency modes (below 250 cm$^{-1}$) in the three spectra is
notable.\\

\noindent
Substituting natural carbon with the $^{13}$C isotope to give a
-S$^{13}$CN label, allows one to probe isotope effect on the IR
spectrum which was done for Ala93S$^{13}$CN in lysozyme. In Figure
\ref{sifig:iso_alascn93}, -SCN and -S$^{13}$CN label-specific IR spectra
(A), IR spectra of Ala93SCN and Ala93S$^{13}$CN residue (B), and full
lysozyme with natural and isotopically substituted -S$^{13}$CN (C) are
reported. The orange dashed line represents IR spectra containing
isotopically -S$^{13}$CN label. In Figure \ref{sifig:iso_alascn93}A, -SCN
and isotope-substituted -S$^{13}$CN label-specific IR spectra are
depicted for the Ala93SCN. Upon comparison, it is noticeable that the
CN-stretching frequency shifts to the red by --47 cm$^{-1}$ for the
heavier -S$^{13}$CN labelled Ala93 residue, consistent with related
findings for -CN labelled metal complexes in aqueous solution, such as
Cu$^{13}$CN, for which the red shift was --42
cm$^{-1}$.\cite{loehr:1992} While the other fundamental peaks of
Ala93SCN are 414 and 690 cm$^{-1}$, those of the S$^{13}$CN-attached
are 406 and 687 cm$^{-1}$, respectively. The isotope effect observed
for these peaks is less pronounced than the shift in the C-N
stretching vibrational mode.\\

\noindent
Next, the -SCN, modified AlaQQSCN, and full protein IR spectra were
determined for all 15 remaining SCN-labelled alanine residues, see
Figure \ref{sifig:ala_3ir}. The -SCN fundamental peaks vary within the
ranges of 2216--2220 cm$^{-1}$ for $v_{1}$, 413--425 cm $^{-1}$ for
$v_{2}$, and 686--713 cm$^{-1}$ for $v_{3}$. However, when comparing
the AlaQQSCN subsystem and full protein spectra, they also differ
depending on the position at which the Ala-residue is located along
the polypeptide chain. Hence, IR spectra of each locally labelled
alanine residue may have unique fingerprints for lysozyme.\\

\begin{figure}[H]
\begin{center} 
\includegraphics[width=1\textwidth]{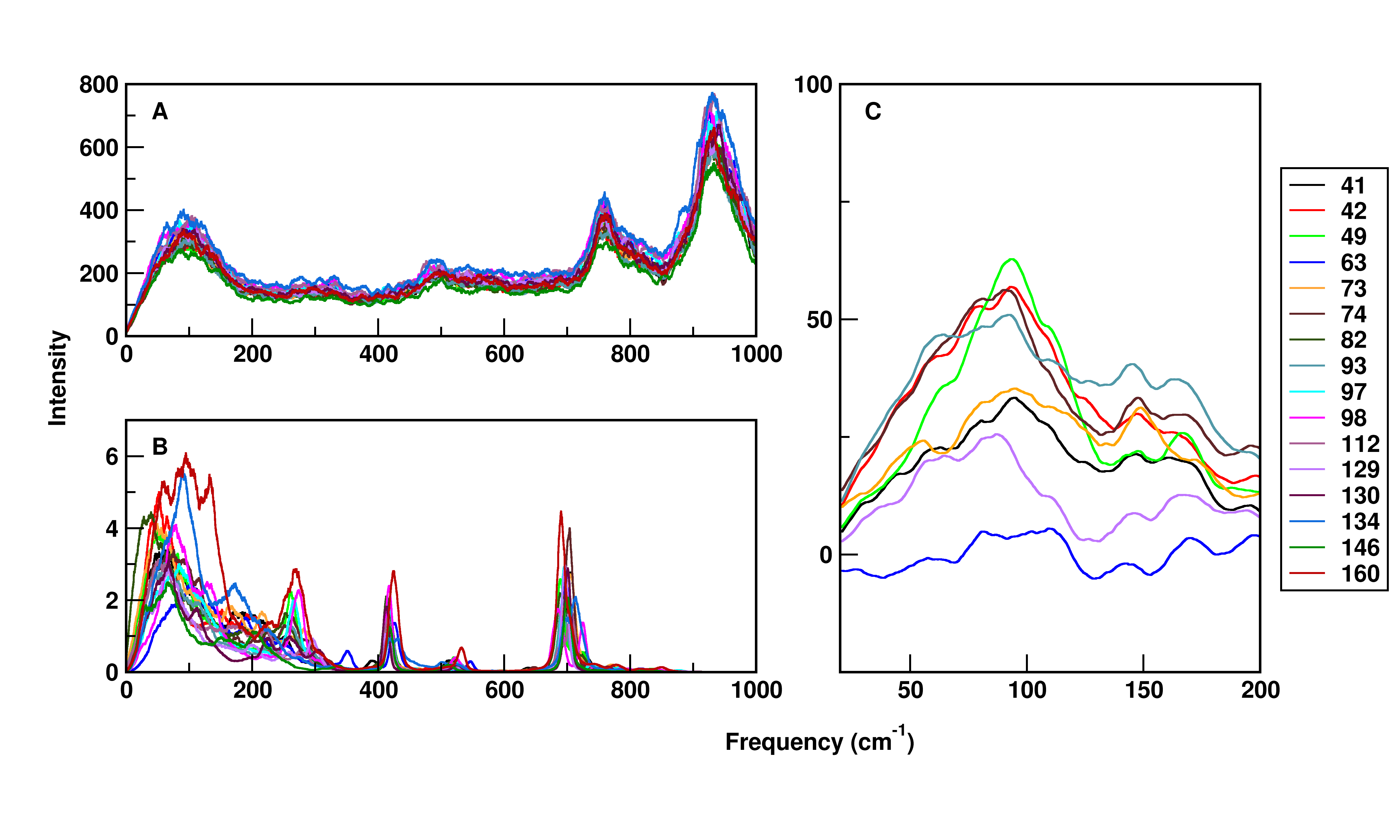}
\caption{Panel A: IR spectra from the total protein dipole moment time
  series; panel B: label-specific IR spectra for all 16 AlaQQSCN
  residues in lysozyme; panel C: the difference IR spectra for the
  entire protein between WT lysozyme and eight selected
  AlaQQSCN-labelled lysozyme variants in the far-infrared part of the
  spectrum.  The difference spectrum for Ala129SCN is discussed in the
  text.}
\label{fig:ir_all}
\end{center}
\end{figure}

\noindent
To further investigate this, the IR spectrum for unlabelled (WT)
lysozyme was also determined, see black trace in Figure
\ref{sifig:ir_org}. Overall, the spectrum for the WT follows that of
all modified proteins. The maximum amplitude in the low-frequency
region (below $\sim 250$ cm$^{-1}$) is found for the Ala134SCN variant
and Ala146SCN has the lowest amplitude in this frequency
interval. Analysis of the low-frequency spectra for full lysozyme
(Figure \ref{fig:ir_all}A) and the label-specific spectra (Figure
\ref{fig:ir_all}B) for all AlaQQSCN residues confirm that the
low-frequency part of the IR spectrum contains label-specific
signatures.  In Figure \ref{fig:ir_all}C, the difference far-infrared
spectra between the unmodified and the SCN-labelled proteins for eight
alanine residues from Figure \ref{fig:ir_all}A are reported. Such
difference spectra change depending on the position of the
SCN-labelled alanine residue along the polypeptide chain. This may
provide further information on the local dynamics.\\ \\

\noindent
It is interesting to note that the low-frequency part of the IR
spectrum was found to be sensitive to the helical content of
proteins.\cite{MS.lf_ss:2017} Towards the red part of the maximum at
100 cm$^{-1}$ the measured spectra superimpose whereas towards the
blue part at 150 cm$^{-1}$ the intensity of lysozyme ($\alpha-$helical
content of 48 \%) is twice that of concanavalin (2
\%).\cite{MS.lf_ss:2017} Hence, it is of interest to consider changes
in this frequency range upon SCN-labelling the
alanine residues. Similarly, far-infrared spectroscopy of peptides,
lysozyme, and other globular proteins established a pronounced peak
between 100 and 200 cm$^{-1}$ depending on the protein
considered.\cite{MS.lf_ss:2017,hellwig:2017,ding:2012,ferraro:1971}\\

\noindent
To link protein structural modifications and changes in the
(far-)infrared spectroscopy, the case of Ala129SCN was considered
specifically. The difference far-infrared spectrum is the violet line
in Figure \ref{fig:ir_all}C. For the unmodified protein, the secondary
structure analysis tool of VMD\cite{VMD96} identifies, among others,
three helices extending across residues 126--134, 137--141, and
143--155. Helices 126--134 and 143--155 are positioned in a parallel
fashion. For the modified Ala129SCN protein it is found that on the 2
ns time scale residues 135--142 are in a loop region. In other words,
introducing the -SCN label at position 129 partially destroys the
helical structure in this region of the protein. This is confirmed by
considering the distance between the C$_{\alpha}$ atoms of residues
Ala129 and Arg154 which increases from $\sim 7.7$ \AA\/ to $\sim 9.3$
\AA\/ after introducing the -SCN label. The difference spectrum between
WT and the Ala129SCN-labelled variant (violet line in Figure
\ref{fig:ir_all}C) features maxima at 88 and 166 cm$^{-1}$.\\

\subsection{Dynamics of the Spectroscopic Probe}
\begin{figure}[H]
\begin{center}
\includegraphics[width=0.3\textwidth,angle=-90]{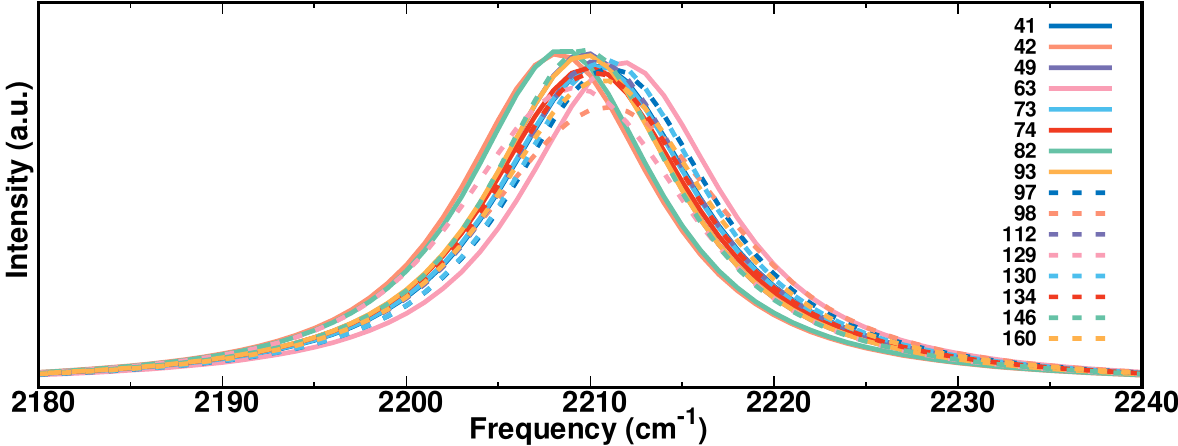}
\caption{1D IR spectra obtained from INM analysis for all 16 AlaQQSCN
  residues in Lysozyme.}
\label{fig:1dir}
\end{center}
\end{figure}

Next, the local dynamics and spectroscopy of the CN-stretch for all
modified AlaQQSCN residues were considered. For this, the FFCFs were
determined for all 16 modified lysozyme variants using INM calculations. 
The 1-dimensional IR spectra in the region
of the -CN stretch for each -SCN label, see Figure \ref{fig:1dir}, can
be calculated from the FFCF. The center frequencies cover a range of
$\sim 4$ cm$^{-1}$, extending from 2208 cm$^{-1}$ for Ala42N$_3$ to
2212 cm$^{-1}$ for Ala63N$_3$ and the full width at half maximum
ranges from 13 to 16 cm$^{-1}$. This is consistent with the analysis
based on the dipole-dipole correlation function (see above) for which
the center frequencies cover the range 2216 to 2220 cm$^{-1}$,
slightly shifted to the blue by $\sim 10$ cm$^{-1}$.\\

\begin{figure}[H]
\begin{center}
\includegraphics[width=1\textwidth]{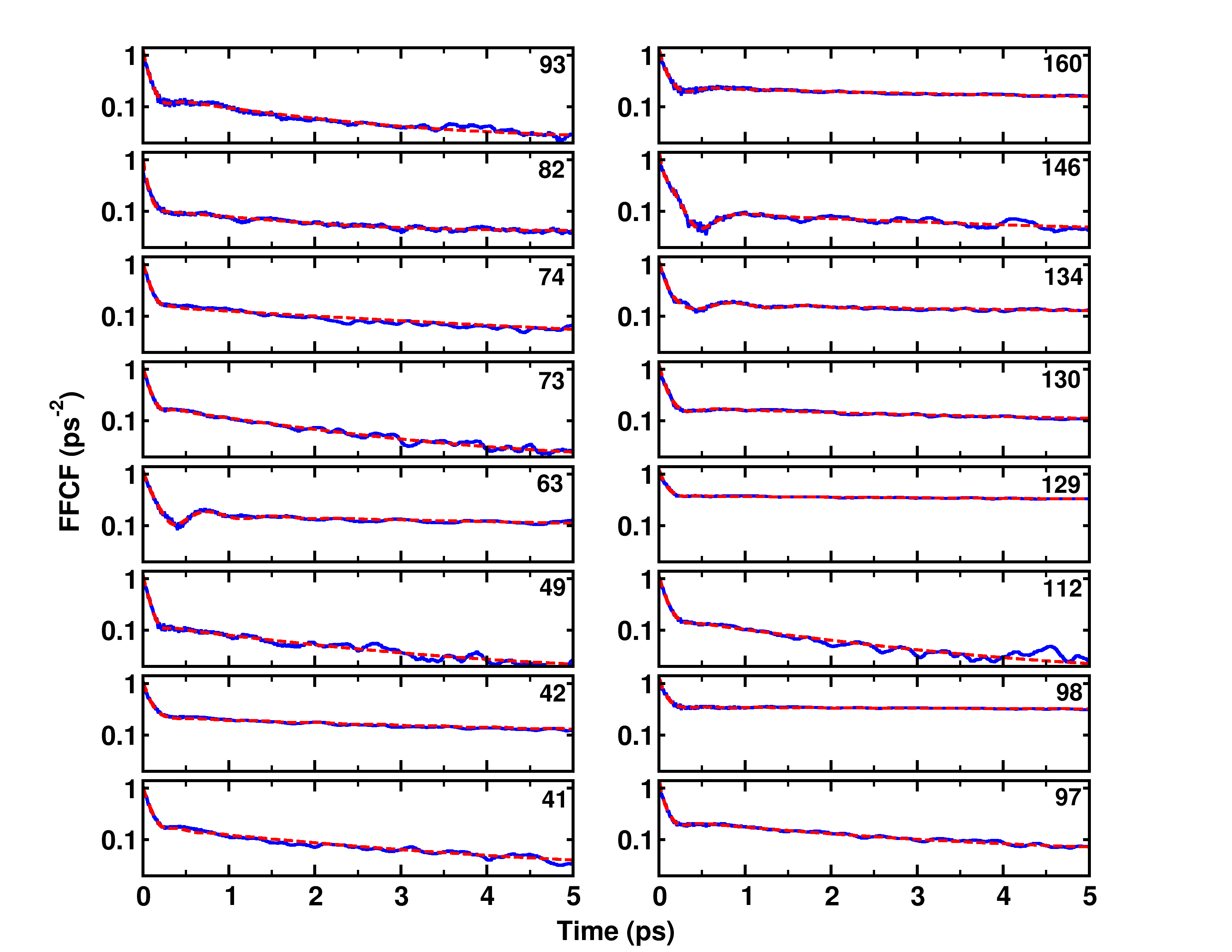}
\caption{FFCFs obtained using INM analysis for all 16 AlaQQSCN. The
  labels in each panel refer to the alanine residue which carries the
  -SCN probe. Blue traces are the raw data and red dashed lines show
  the fits to Eq. \ref{eq:ffcffit}. The $y-$axis is logarithmic.}
\label{fig:ffcf}
\end{center}
\end{figure}

\noindent
Next, the FFCFs themselves are considered, see Figure \ref{fig:ffcf}
and Table \ref{sitab:ffcffit}. The majority of FFCFs have a
comparatively ``simple'' behaviour in that they show a rapid initial
decay on the sub-ps time scale, followed by a more or less extended
decay towards zero on the 5 to 10 ps time scale. Modification sites
which differ from this are Ala63SCN, Ala134SCN, and Ala146SCN which
show a more pronounced rebound during the first picosecond. As can be
seen from Figure \ref{fig:ffcf}, residues
Ala82SCN, Ala98SCN and Ala129SCN exhibit a considerably slower decay
$\tau_2$, which may be attributed to the environmentally crowded
positioning of the residues. For example, Ala98SCN is located within
the interior of the $\alpha$-helix, with another helical structure
present nearby on the residue's vicinity. Ala129SCN is positioned in
the middle of the $\alpha$-helix and, in addition, different
$\alpha$-helices are located on both sides of it. Ala130SCN and
Ala134SCN are also located within the same helical structure as
Ala129SCN, but they are positioned further away from the crowded side
of the $\alpha$-helix compared to Ala129SCN. Therefore, the decay
times are faster ($\sim 23.3$ ps compared with 4.5 and 6.9 ps,
respectively). The two decay times are on the sub-ps ($\tau_1 \leq
0.07$ ps) and several picosecond time scales $1.47 \leq \tau_2 \leq
23.30$ ps and the amplitude of the fast process is invariably larger
than that of the slow process.\\

\noindent
The minimum in the FFCF at early correlation times has been observed
in experiments\cite{cheatum:2016} and from
simulations\cite{hynes:2004,skinner:2006,MM.cn:2013,MM.lys:2021} and
has been related to the strength of the intermolecular interaction
between the IR probe and its environment.\cite{hynes:2004,MM.cn:2013}
Such minima were found for residues Ala63SCN, Ala134SCN, and
Ala146SCN. However, they are considerably less pronounced than those
observed for azide labelled (-N$_3$) in lysozyme.\cite{MM.lys:2021}\\

\noindent
Residues Ala98SCN, Ala129SCN, and Ala160SCN feature an asymptotic
value $\Delta_0^2$ (``static component'') larger than 0.1
cm$^{-1}$. Such finite values of $\Delta_0^2$ have been associated
with residual dynamics which has not decayed on the time scale of the
analysis. A value of $\Delta_0^2 = 0.25$ ps$^{-2}$, i.e. $\Delta_0^2 =
0.5$ ps$^{-1}$ corresponds to $\sim 2.5$ cm$^{-1}$. This is consistent
with experimentally measured static components for a nitrile probe
(cyanophenylalanine) in HP35 (2.2 cm$^{-1}$) and in S-peptide (2.4
cm$^{-1}$)\cite{Chung.hp352dir.pnas.2011,Bagchi.cnprot2dir.jpcb.2012}. Furthermore,
static inhomogeneous components were observed in the spectral
diffusion of azide probes in a peptide bound to a
protein.\cite{hamm.jpcb.2012} Finally, previous simulations of
cyano-substituted phenol in lysozyme reported static components of
$\sim 2$ cm$^{-1}$, comparable with the present
findings.\cite{MM.lys:2017}\\

\begin{table}[H]
\footnotesize
%\scriptsize
\centering
\caption{Parameters obtained from fitting the normalized FFCF
  to Eq. \ref{eq:ffcffit} for INM frequencies for all different AlaQQSCN
  residues in lysozyme. Average frequency $\langle\omega\rangle$ of
  the asymmetric CN-stretch in cm$^{-1}$, the amplitudes $a_1$ and
  $a_2$ in ps$^{-2}$, the decay times $\tau_1$ and $\tau_2$ in ps, the
  parameter $\gamma$ in ps$^{-1}$ and the offset $\Delta_0^2$ in
  ps$^{-2}$.}  \centering
\begin{tabular}{|r|c|crc|cc|c|}
\hline\hline
Res& $\langle\omega\rangle$ & $a_{1}$ & $\gamma$ &$\tau_{1}$ &$a_{2}$ &$\tau_{2}$ & $\Delta_0^2$\\
\hline
\textbf{41} & 2209.39   & 0.84 &  14.38   & 0.07   & 0.16   & 2.21  &  0.024    \\
\hline
\textbf{42} & 2207.26   &  0.86  & 9.13   & 0.07   & 0.14  & 4.67   & 0.083   \\
\hline
\textbf{49} & 2208.73   &  0.89  & 13.11   & 0.06   & 0.11 & 1.82   & 0.015   \\
\hline
\textbf{63} & 2210.69  & 0.88 & 8.05   & 0.07   & 0.1   & 8.52  & 0.042   \\
\hline
\textbf{73} & 2209.33   & 0.82 & 13.29  & 0.06   & 0.18   & 1.58   & 0.017   \\
\hline
\textbf{74} & 2208.94   & 0.86 & 16.72   & 0.07   & 0.14   & 3.82   & 0.018   \\
\hline
\textbf{82} & 2207.51  & 0.94 & 0.58   & 0.04   & 0.06   & 20.72   & 0.000   \\
\hline
\textbf{93} & 2208.71   & 0.86 & 9.35  & 0.04   & 0.14 & 1.47   & 0.024   \\
\hline
\textbf{97} & 2210.09   & 0.81  & 9.11  & 0.03   & 0.19   & 2.18   &  0.053  \\
\hline
\textbf{98} & 2209.91  &0.91 & 10.88 & 0.06   &0.09    & 15.98  & 0.256   \\
\hline
\textbf{112} & 2209.42   & 0.84 & 9.08   & 0.06   & 0.16   & 1.72   & 0.014   \\
\hline
\textbf{129} & 2208.13   & 0.75 & 12.08   & 0.07   & 0.25   & 23.30  & 0.129   \\
\hline
\textbf{130} & 2209.93   & 0.89 & 7.87   & 0.05   & 0.11   & 4.54 &  0.077  \\
\hline
\textbf{134} &2209.22  & 0.92 & 7.19   & 0.06   & 0.08   & 6.91   & 0.087   \\
\hline
\textbf{146} & 2208.67  & 0.93 & 5.18   &0.03    &0.07    & 3.37   & 0.034   \\
\hline
\textbf{160} & 2109.58  & 0.87 & 8.36  & 0.05   & 0.13   & 4.94   & 0.111  \\
\hline\hline
\end{tabular}
\label{sitab:ffcffit}
\end{table}

\section{Discussion and Conclusion}
The present work investigated and quantified changes in spectroscopy
and local dynamics for SCN-labelled alanine residues in lysozyme. From
two different analyses of the IR spectroscopy in the region of the
CN-stretch vibration it was found that the maxima in the 1-dimensional IR spectra
cover a range of 4 cm$^{-1}$. The dynamics as inferred from the decay
of the FFCF for the 16 sites is characterized by a first sub-ps decay
time and a slower time scale $\tau_2$ which ranges from 1.5 ps to 23.3
ps. Also, several positions along the polypeptide chain feature a
static component $\Delta_0^2$ different from zero which points towards
arrested dynamics on the time scale of the analysis. The
label-specific difference spectra between unmodified and modified
lysozyme indicate that the low frequency modes respond differentially
to the position at which the label is placed.\\

\noindent
Other spectroscopic probes have been considered in the past to
elucidate protein structural dynamics, including nitrile probes to
clarify the role of electrostatic fields in enzymatic
reactions\cite{boxer:2014,hammes-schiffer:2013} or to elucidate the
mode of drug binding to proteins.\cite{hochstrasser:2013,blasie:2009}
Other labels include cyanamide,\cite{cho:2018} sulfhydryl vibrations
of cysteines,\cite{hamm:2008} deuterated carbons,\cite{romesberg:2011}
carbonyl vibrations of
metal-carbonyls,\cite{kevin:2012,kevin.2:2012,kevin:2014} or
cyanophenylalanine.\cite{thielges:2019} Finally, for azidohomoalanine
experiments\cite{hamm:2012} and simulations\cite{MM.lys:2021} have
explored its utility as a position-sensitive probe. When attached to
lysozyme, -N$_3$ was found to cover a frequency range of $\sim 15$
cm$^{-1}$ compared with a frequency span of $\sim 10$ cm$^{-1}$ when
attached to Val, Ala, or Glu in PDZ2.\cite{hamm:2012} This compares
with 4 cm$^{-1}$ in the present case for -SCN and may be in part due
to using a simpler point charge-based model rather than a 
more elaborate multipolar model for the
electrostatics.\cite{MM.mtp:2012,MM.mtp:2013,MM.dcm:2014,MM.mdcm:2017}\\

\begin{figure}[H]
\centering
\includegraphics[width=1\textwidth,angle=0]{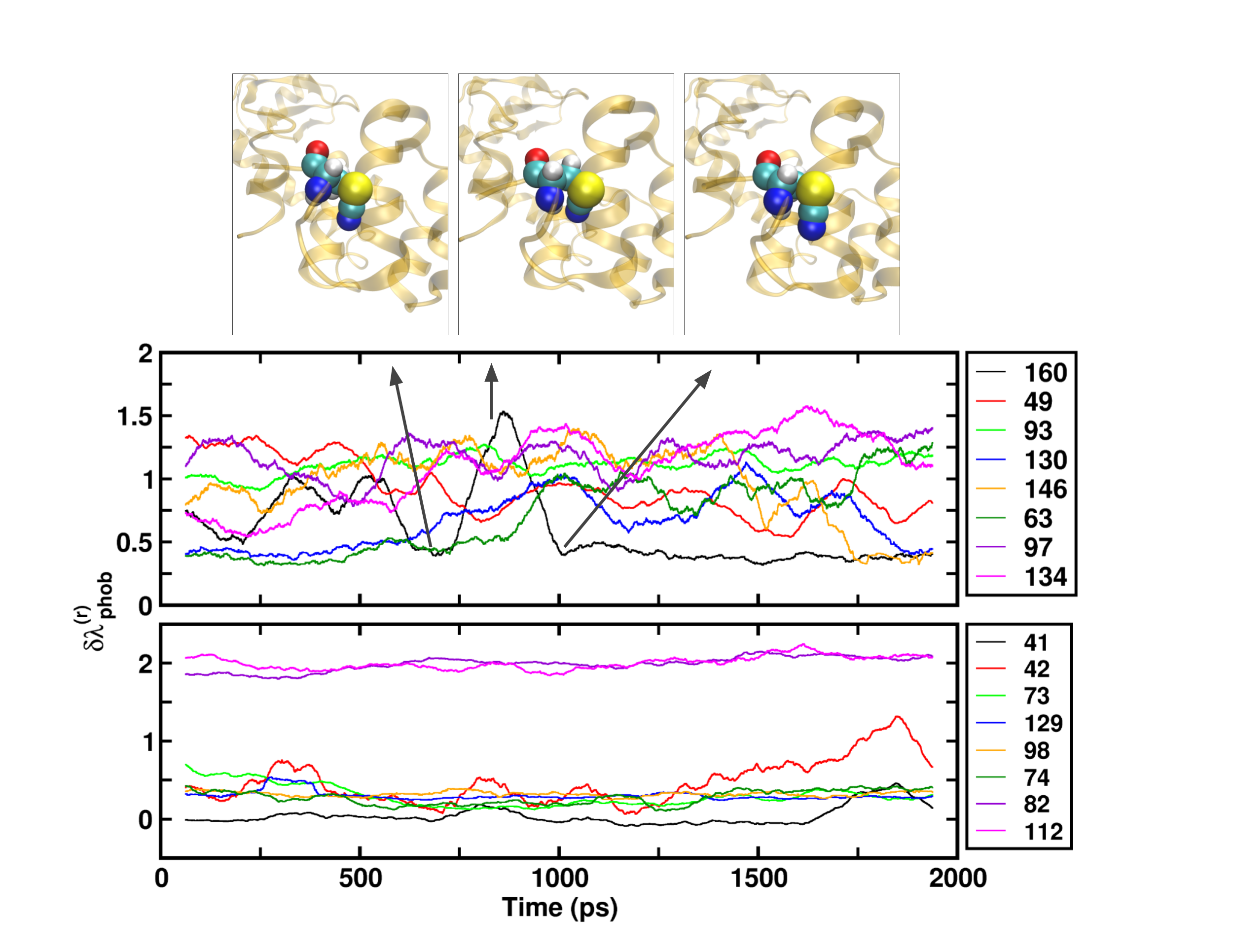}
\caption{Local hydrophobicity as a function of simulation time for all
  alanine residues for SCN-labelled lysozyme. The LH coefficient was
  determined from Eq. \ref{eq:lh}. Values for $\delta \lambda_{\rm
    phob} \sim 0$ and $\delta \lambda_{\rm phob} \sim 2$ are
  indicative of hydrophobic and hydrophilic environments,
  respectively.\cite{Willard-jpcb-2018,MM.hb:2020} The top three panels display various structures from MD simulations of Ala160SCN. These structures depict the temporal structural changes in time and are generated around at 630 ps, 840 ps, and 1 ns, respectively.}
\label{fig:hyd}
\end{figure}

\noindent
The FFCFs of only a few residues show recurrences. These features were
found in previous simulations\cite{skinner:2006,MM.cn:2013} and in
experiments\cite{cheatum:2016} and have been related to the strength
of the intermolecular interaction between the IR probe and its
environment.\cite{hynes:2004,MM.cn:2013} Together with the smaller
frequency range and frequency shifts of the CN-stretch vibration this
indicates that the -SCN label interacts less strongly with its
environment. This is consistent with rather narrow frequency shifts
for compounds such as MeSCN in polar, non-polar, and protic solvents,
covering only 9 cm$^{-1}$ for the frequency
maximum.\cite{baiz:2024,gao:2022} This compares with 4 cm$^{-1}$ from
the present work for -SCN attached to protein-alanine residues in
water. The frequency maxima in the present work are around 2220
cm$^{-1}$ whereas experimentally this is shifted to the red by --60
cm$^{-1}$. This is mainly due to the level of theory which was used
for constructing the 3-dimensional PES for the -SCN label. It is
possible that the frequency spread and position of the frequency
maximum shift somewhat more if a more elaborate multipolar or
distributed charge model is
used.\cite{cho:2008,bereau2013,MM.dcm:2014,MM.mdcm:2017} However,
given that the -SCN label is electrically close to neutral,
i.e. SCN$^{0}$, such effects are expected to be small.\\

\noindent
The frequency shifts observed for the 16 different positions at which
the -SCN label was placed is due to both, the exposure of the
spectroscopic label to water and the immediate protein environment.
One way to analyze the solvent exposure of the labelled
alanine-residues is to determine their local hydrophobicity. Figure
\ref{fig:hyd} reports the local hydrophobicity as a function of
simulation time for all 16 SCN-labelled alanine residues in lysozyme.
suggesting the time-dependence characteristic of LH which is more
pronounced for Ala160. On the other hand, without the spectroscopic
label, Ala41 with $\delta \lambda_\mathrm{phob}^{(r)}(t) \approx 0$ is
clearly hydrophobic while Ala82 and Ala112 with $\delta
\lambda_\mathrm{phob}^{(r)}(t) \approx 2$ have high hydrophobicity
character. The LH for alanine residues show both low and high values,
indicative of hydrophobic and hydrophilic environments,
respectively.\\

\noindent
To link protein dynamics and local hydrophobicity the case of
Ala160SCN is considered (black thick line in Figure \ref{fig:hyd} top
panel). Between 600 ps and 1 ns the value of $\delta
\lambda_\mathrm{phob}^{(r)}(t)$ changes from 0.5 to 1.5 and back to
0.5. The corresponding protein structures are shown on top of Figure
\ref{fig:hyd}. It is found the local hydrophobicity at position Ala160
can be directly linked to a structural change in that the orientation
of the -SCN label changes. This modifies the exposure to solvent water
which in turn affects the value of $\delta
\lambda_\mathrm{phob}^{(r)}(t)$. Protein modification has also been
found to influence the local dynamics and hydration for
S-nitrolsylation in myoglobin, hemoglobin, and
K-RAS.\cite{MM.sno:2021,MM.sno:2023}\\

\noindent
In summary, the present work characterized the local dynamics around
SCN-labelled alanine residues in lysozyme. The IR spectroscopy
in the region of the CN-stretch vibration is position-sensitive with
maxima in the 1-dimensional lineshape extending across 4 cm$^{-1}$
which can be detected experimentally and assigned from
simulations.\cite{Suydam.halr.sci.2006,MM.lys:2017} The IR
spectroscopy from the dipole-dipole correlation function and the
analysis of the FFCF are consistent with one another and the FFCFs
exhibit static components depending on the position of the -SCN
label that are supported by earlier experiments and
simulations. Finally, the low-frequency spectra, up to $\sim 300$
cm$^{-1}$, of the labelled proteins differ. This provides a viable
route for characterizing different global motions in the modified
proteins depending on the modification sites.\\

\section*{Acknowledgments}
The authors gratefully acknowledge financial support from the Swiss
National Science Foundation through grant 200021-117810 and to the
NCCR-MUST.\\

\section*{Data Availability Statement}
The data that support the findings of this study are available from
the corresponding author upon reasonable request.

\clearpage

\renewcommand{\thetable}{S\arabic{table}}
\renewcommand{\thefigure}{S\arabic{figure}}
\renewcommand{\thesection}{S\arabic{section}}
\renewcommand{\d}{\text{d}}
\setcounter{figure}{0}  
\setcounter{section}{0}  

\noindent
{\bf Supporting Information: SCN as a Local Probe of Structural Dynamics}\\

\begin{figure}
\centering
\begin{minipage}{.5\textwidth}
  \centering
  \includegraphics[width=1.1\linewidth]{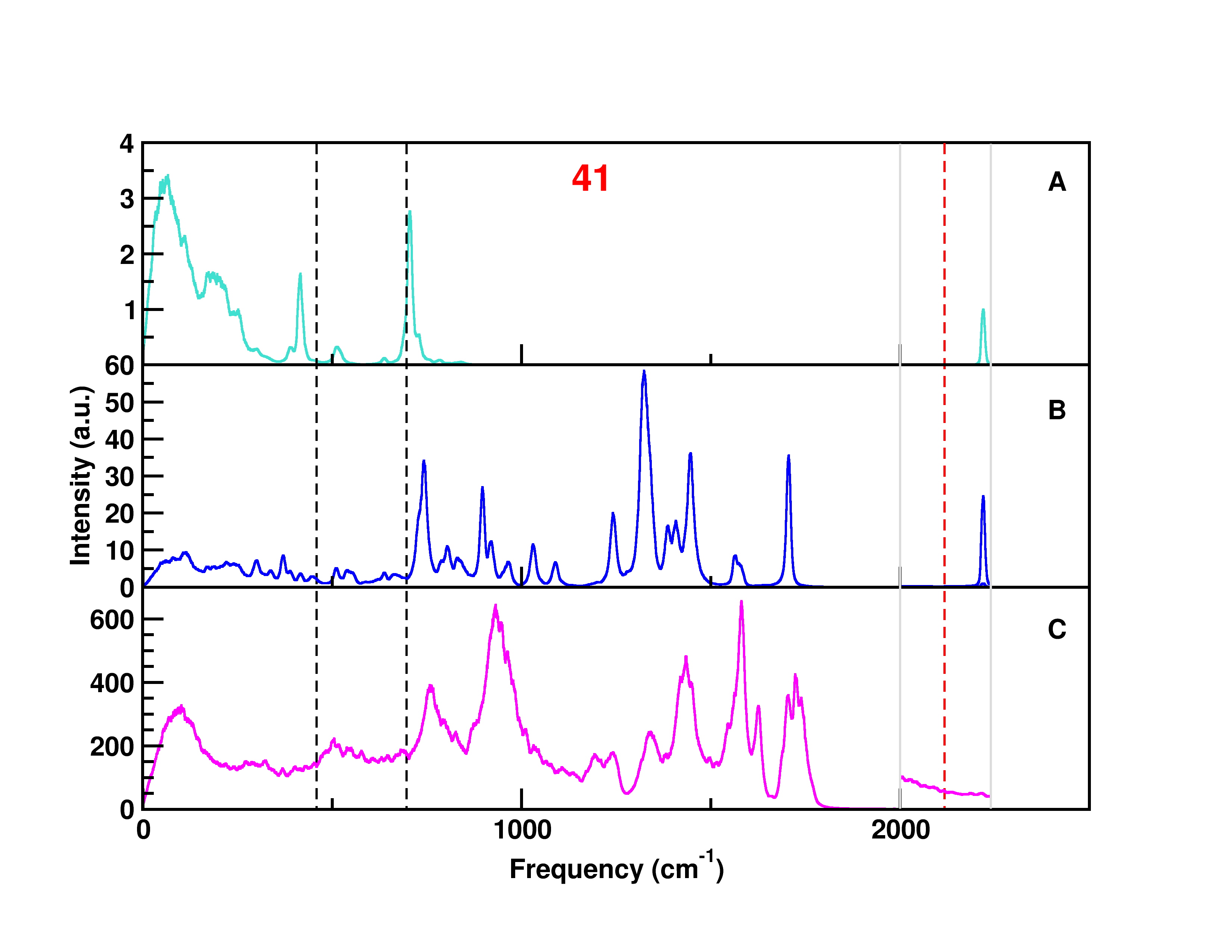}
  \label{fig:ala41}
\end{minipage}%
\begin{minipage}{.5\textwidth}
  \centering
  \includegraphics[width=1.1\linewidth]{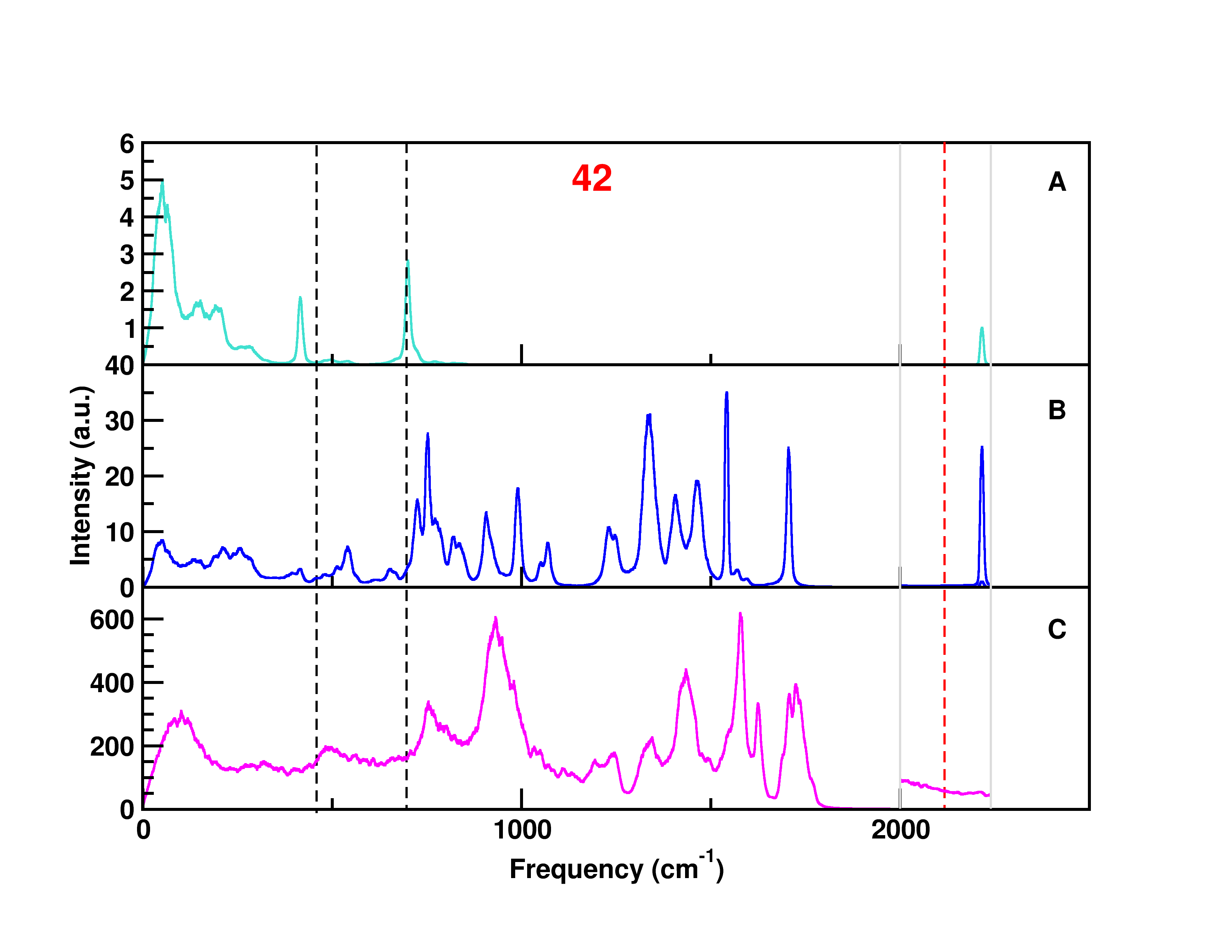}
  \label{fig:ala42}
\end{minipage}
\end{figure}
\begin{figure}
\begin{minipage}{.5\textwidth}
  \centering
  \includegraphics[width=1.1\linewidth]{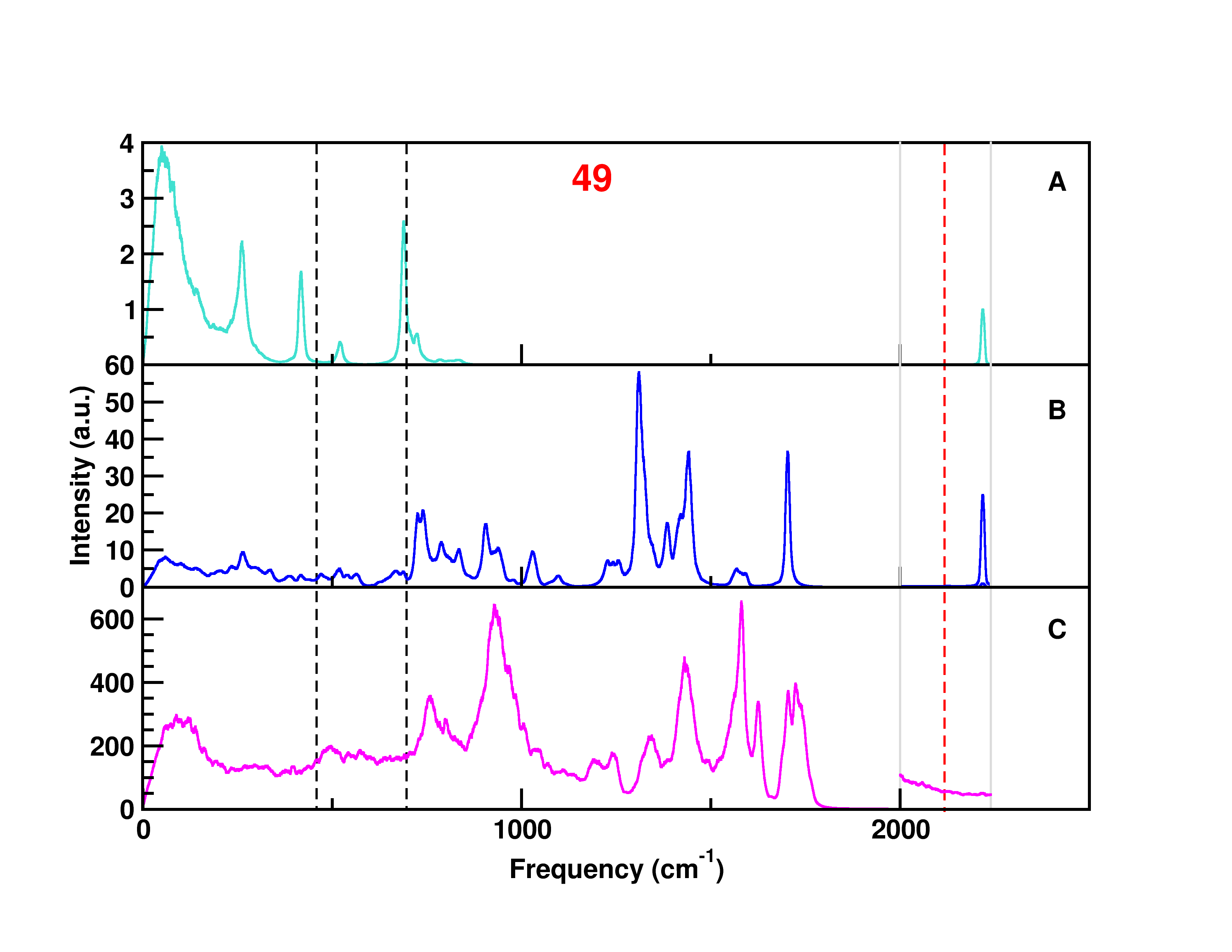}
  \label{fig:ala49}
\end{minipage}%
\begin{minipage}{.5\textwidth}
  \centering
  \includegraphics[width=1.1\linewidth]{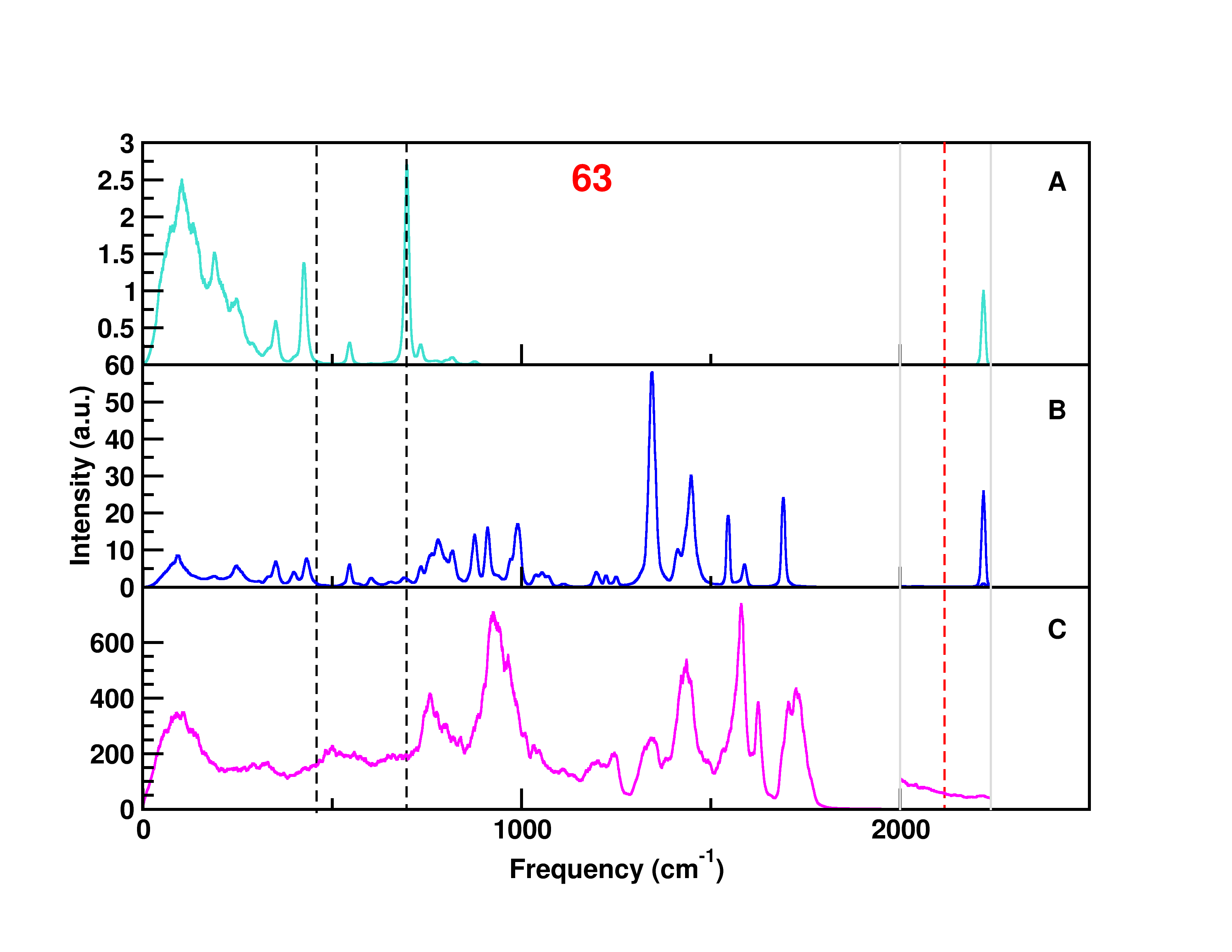}
  \label{fig:ala63}
\end{minipage}
\begin{minipage}{.5\textwidth}
  \centering
  \includegraphics[width=1.1\linewidth]{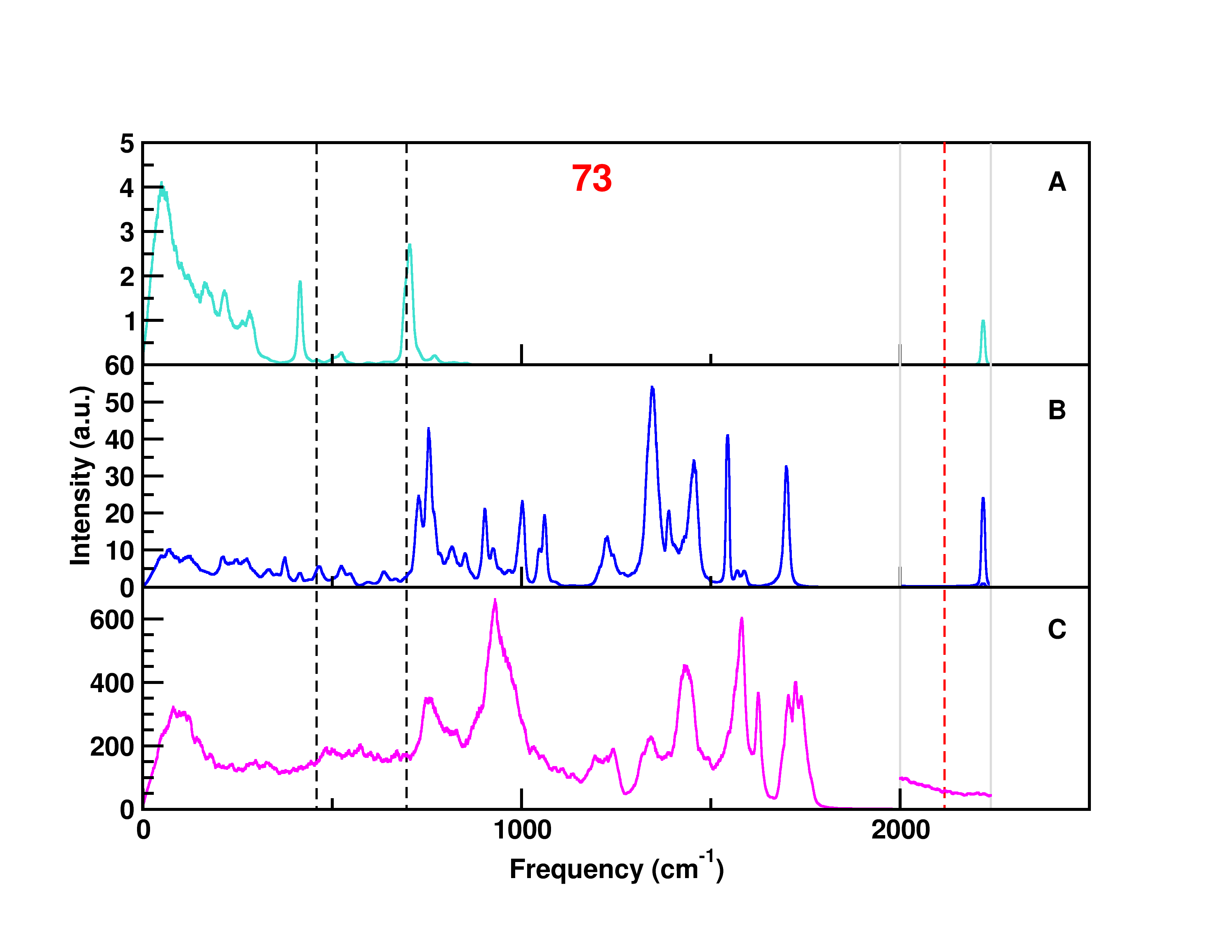}
  \label{fig:ala73}
\end{minipage}%
\begin{minipage}{.5\textwidth}
  \centering
  \includegraphics[width=1.1\linewidth]{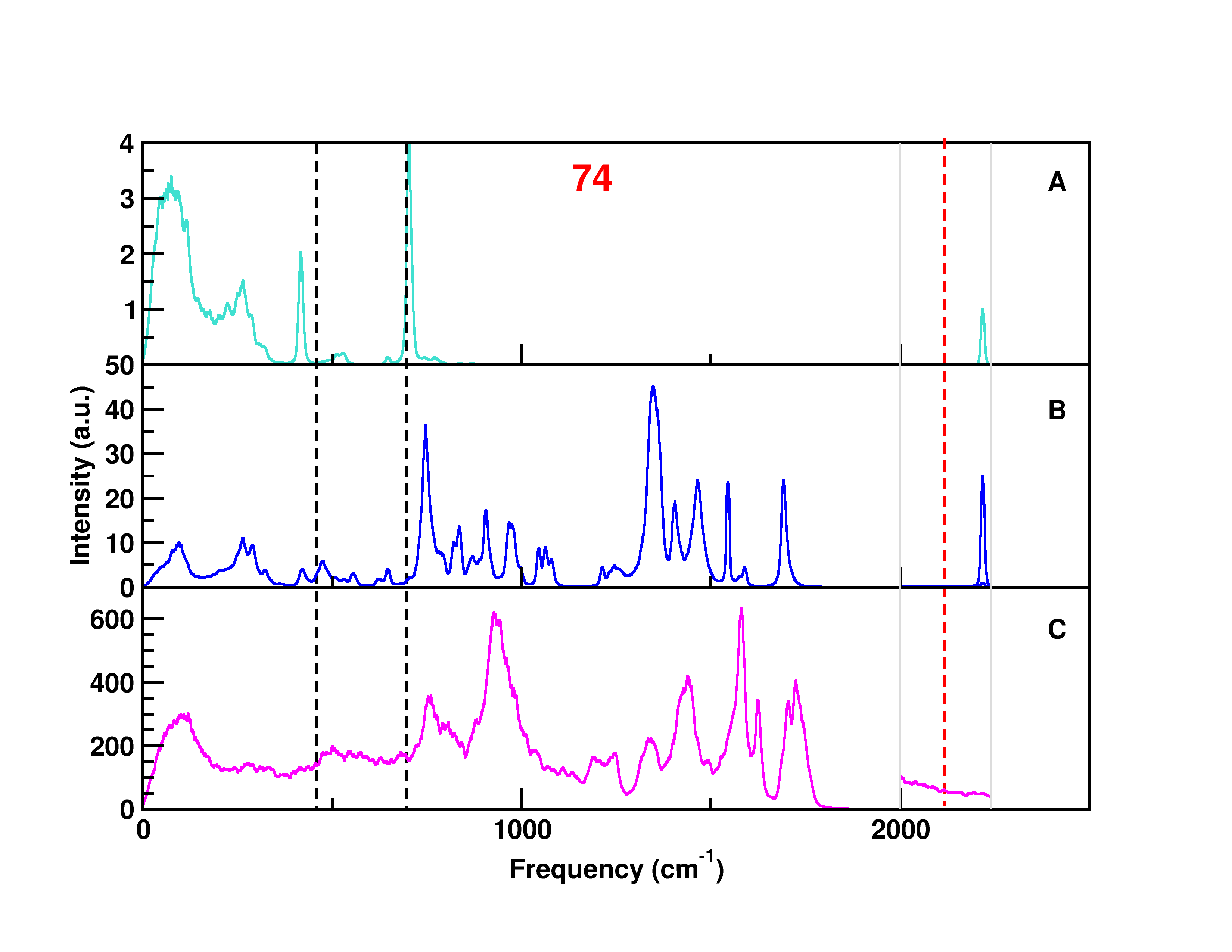}
  \label{fig:ala74}
\end{minipage}
\begin{minipage}{.5\textwidth}
  \centering
  \includegraphics[width=1.1\linewidth]{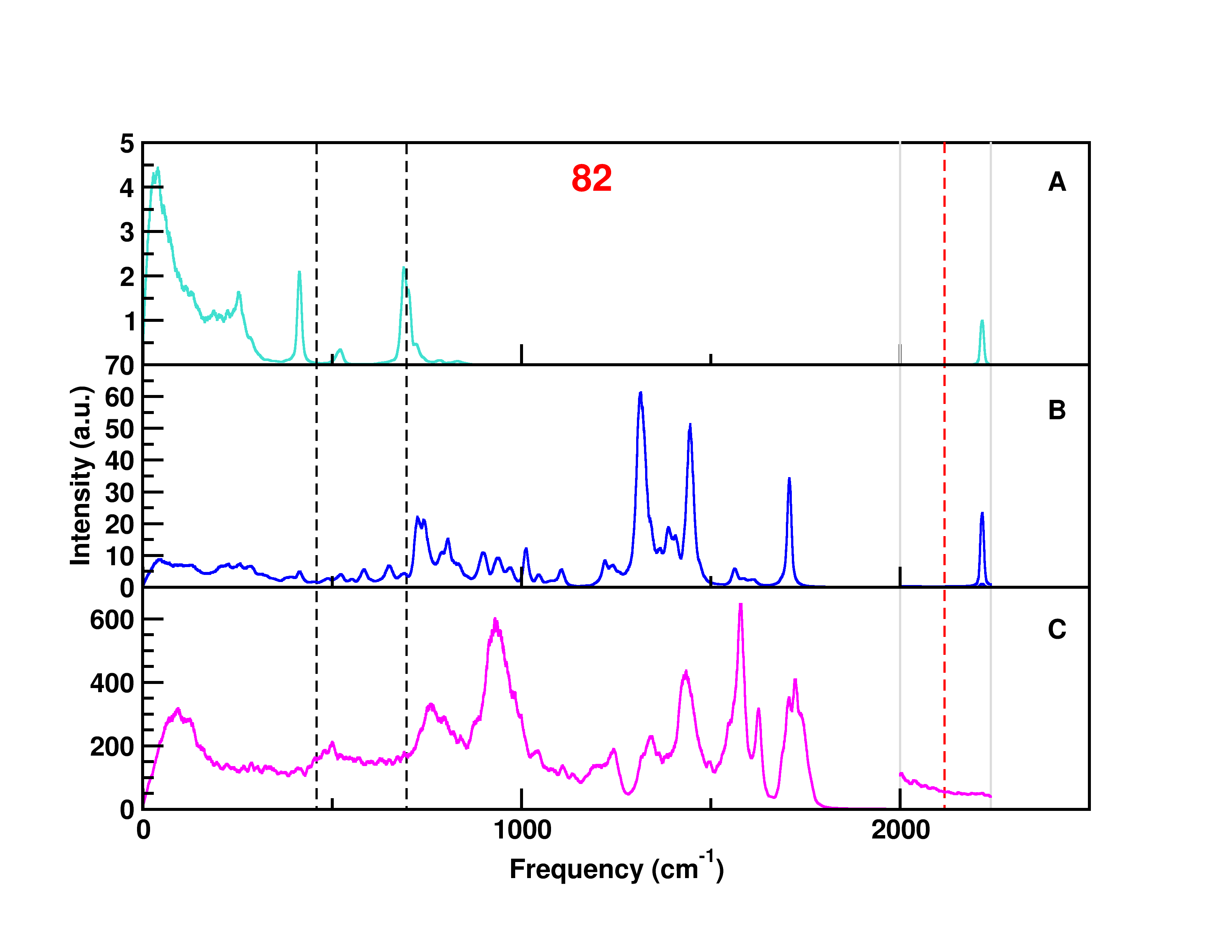}
  \label{fig:ala82}
\end{minipage}%
\begin{minipage}{.5\textwidth}
  \centering
  \includegraphics[width=1.1\linewidth]{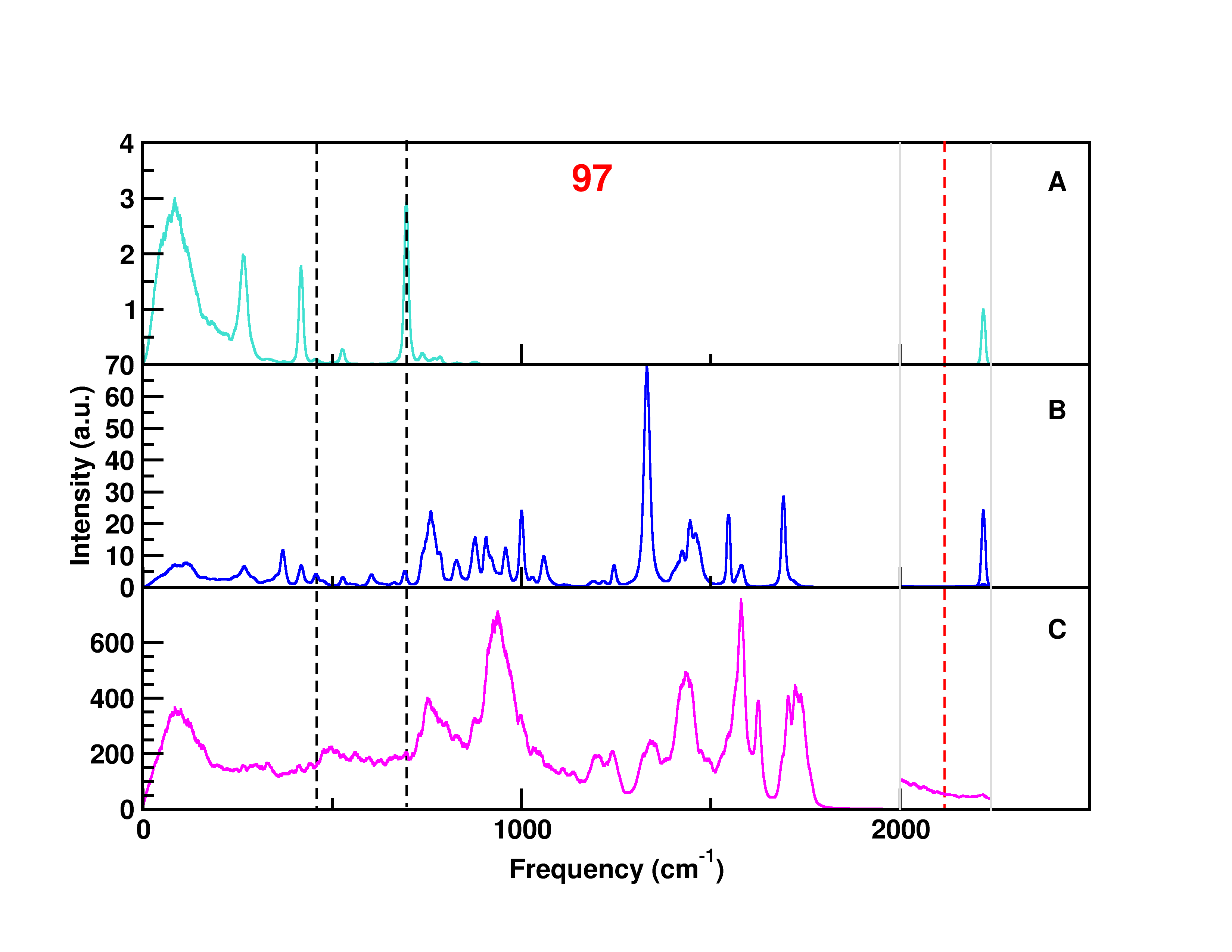}
  \label{fig:ala97}
\end{minipage}
\end{figure}
\begin{figure}
\begin{minipage}{.5\textwidth}
  \centering
  \includegraphics[width=1.1\linewidth]{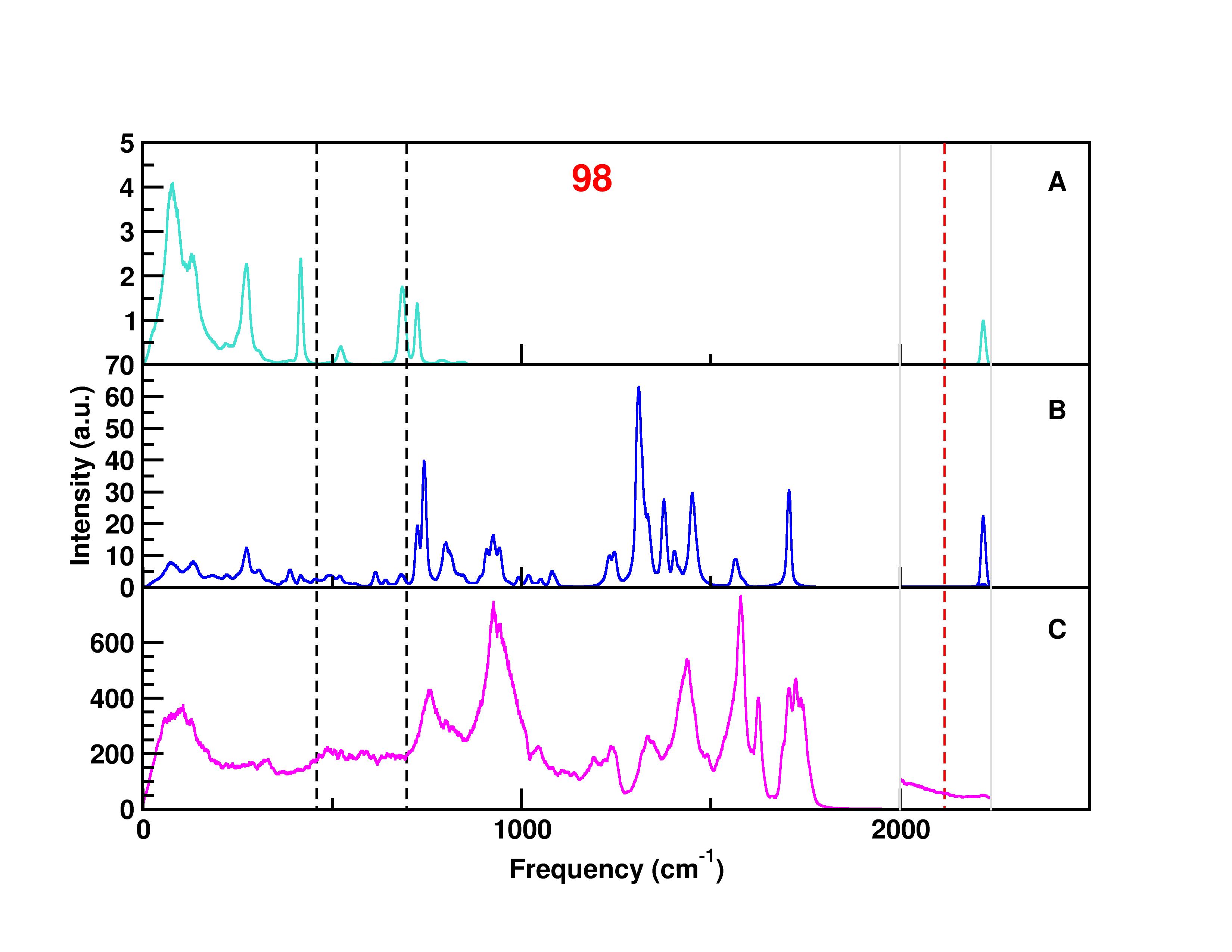}
  \label{fig:ala98}
\end{minipage}%
\begin{minipage}{.5\textwidth}
  \centering
  \includegraphics[width=1.1\linewidth]{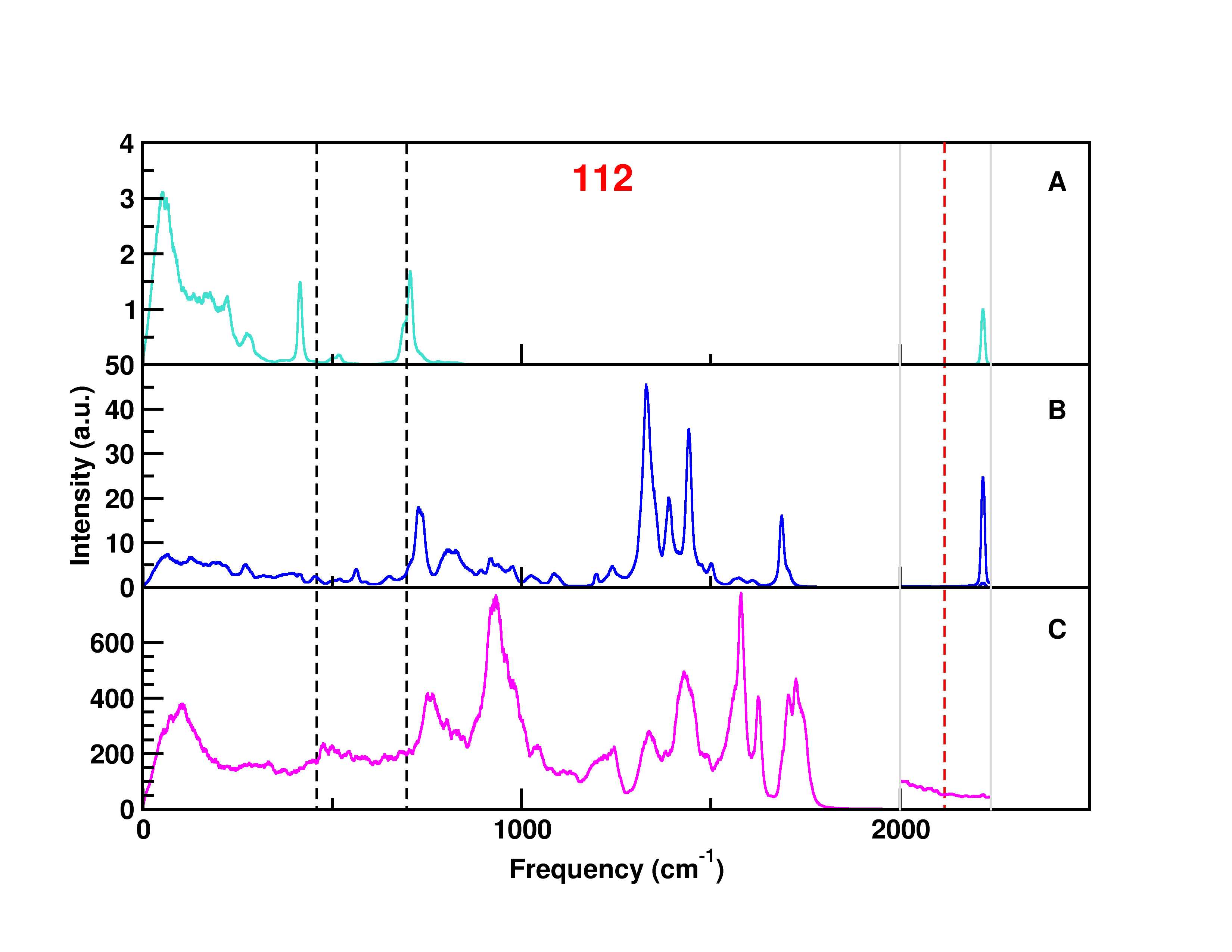}
  \label{fig:ala112}
\end{minipage}
\begin{minipage}{.5\textwidth}
  \centering
  \includegraphics[width=1.1\linewidth]{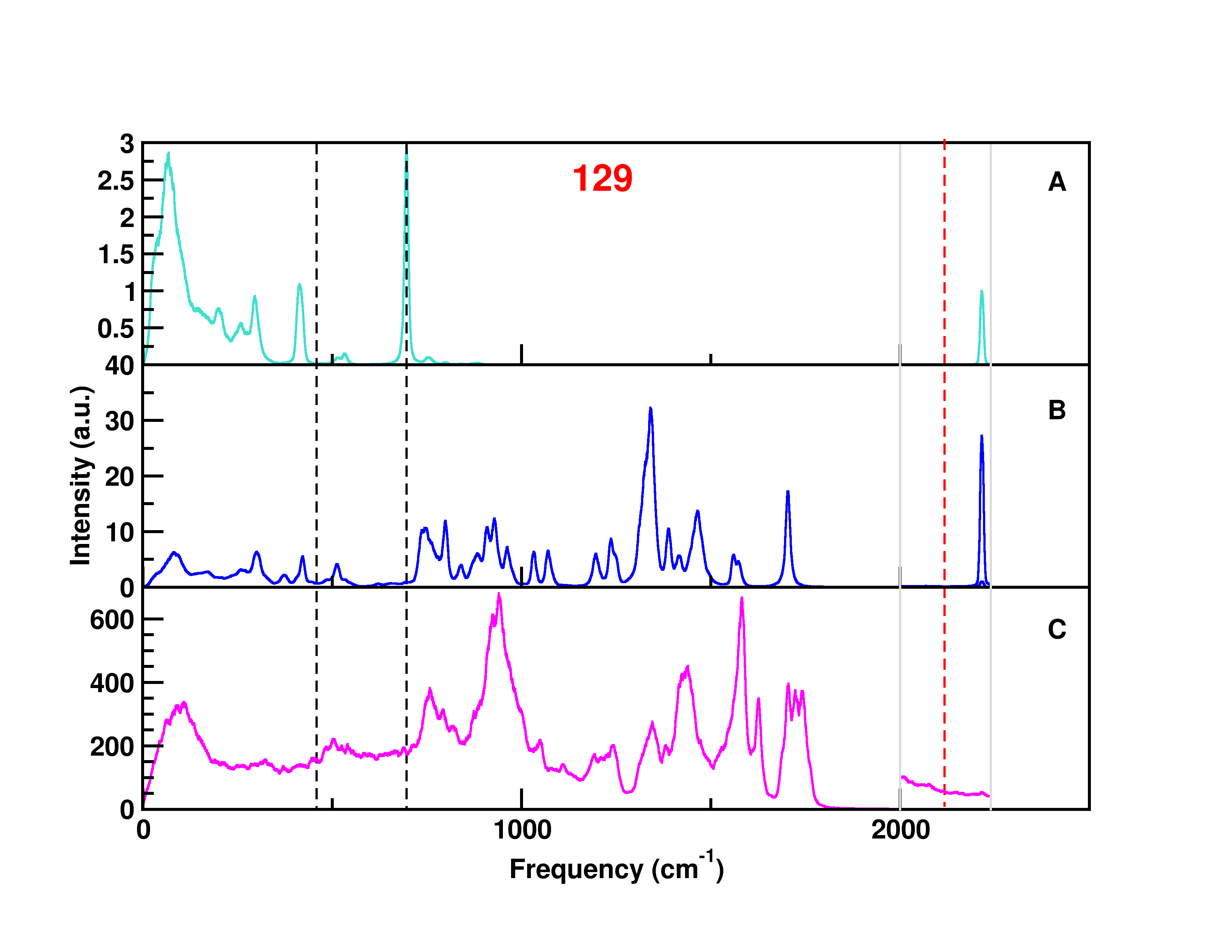}
  \label{fig:ala129}
\end{minipage}%
\begin{minipage}{.5\textwidth}
  \centering
  \includegraphics[width=1.1\linewidth]{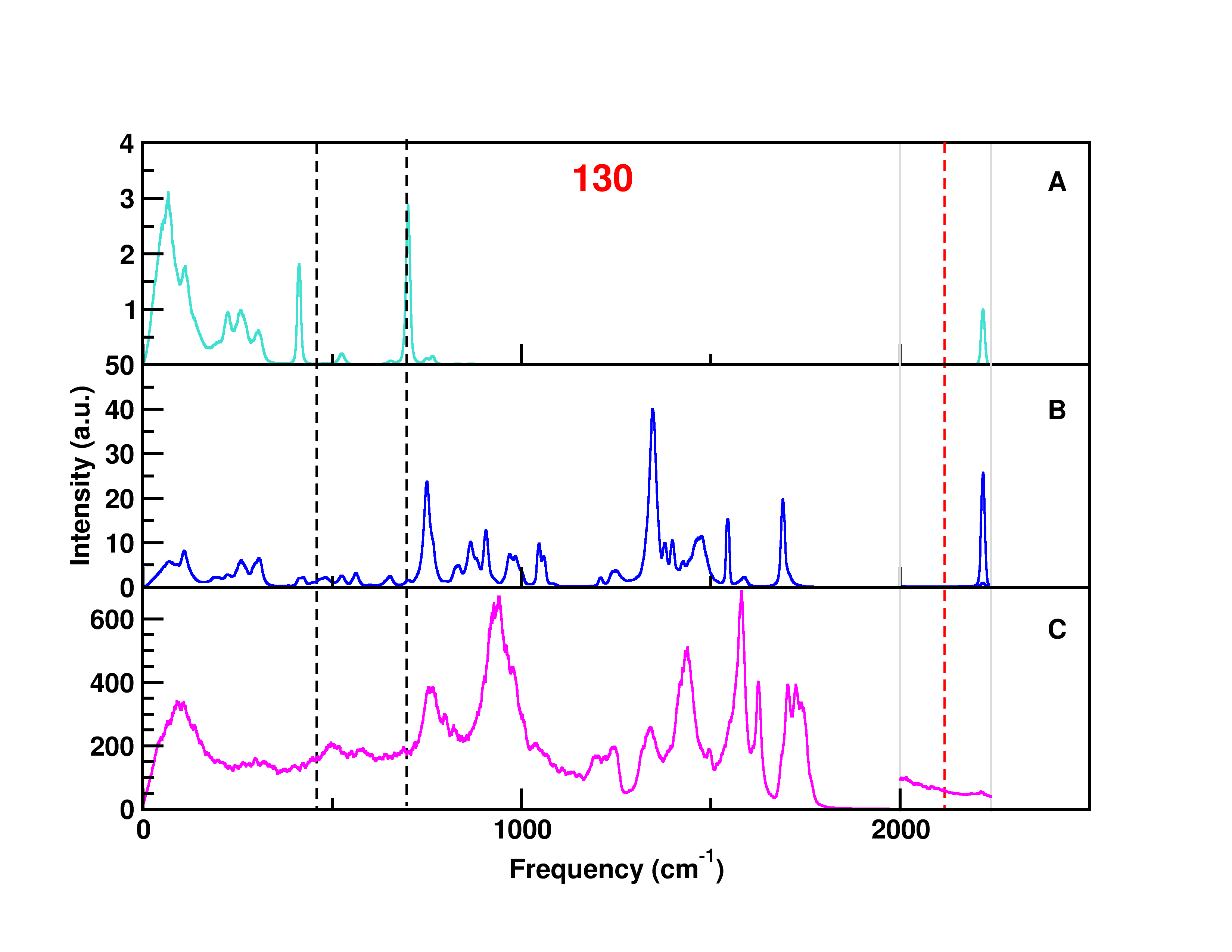}
  \label{fig:ala130}
\end{minipage}
\begin{minipage}{.5\textwidth}
  \centering
  \includegraphics[width=1.1\linewidth]{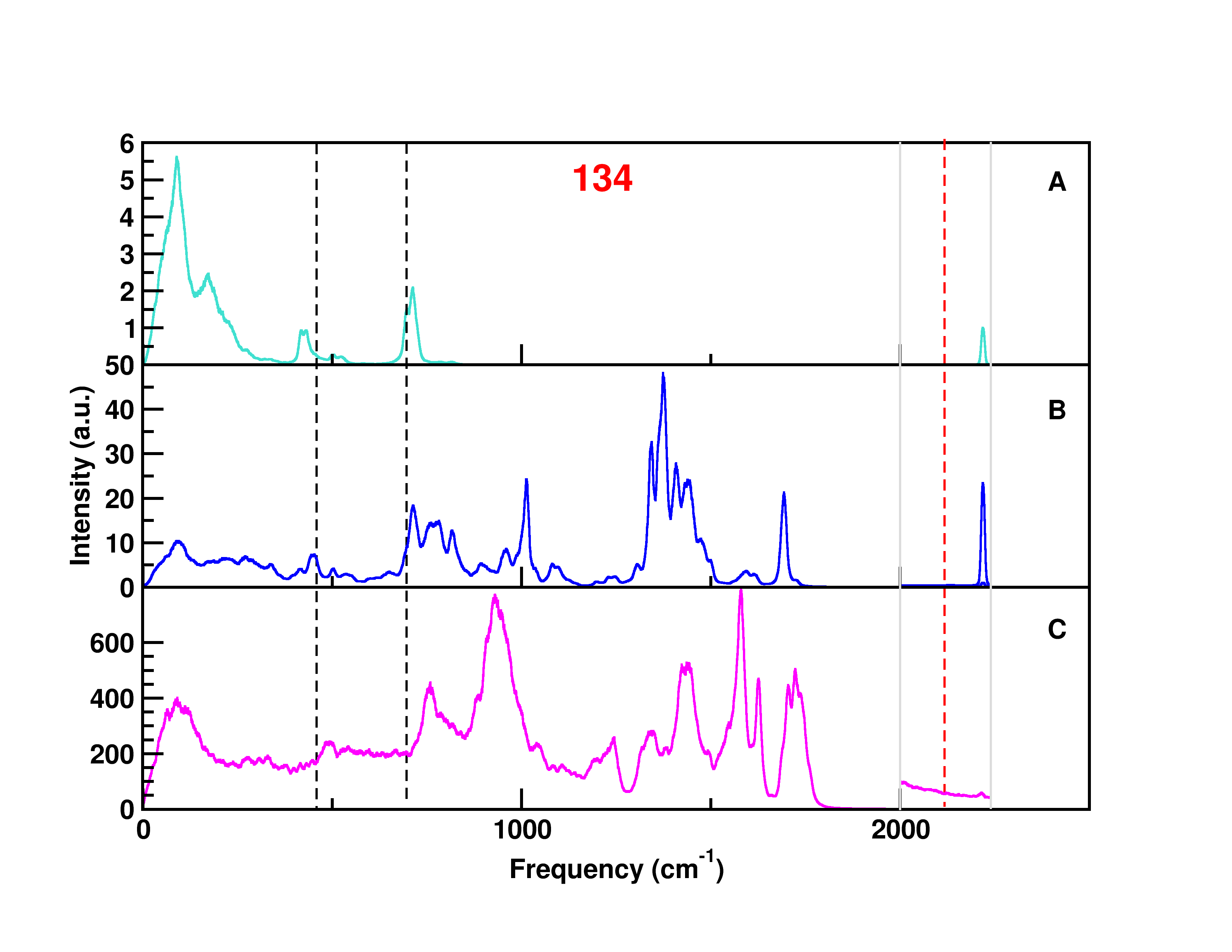}
  \label{fig:ala134}
\end{minipage}%
\begin{minipage}{.5\textwidth}
  \centering
  \includegraphics[width=1.1\linewidth]{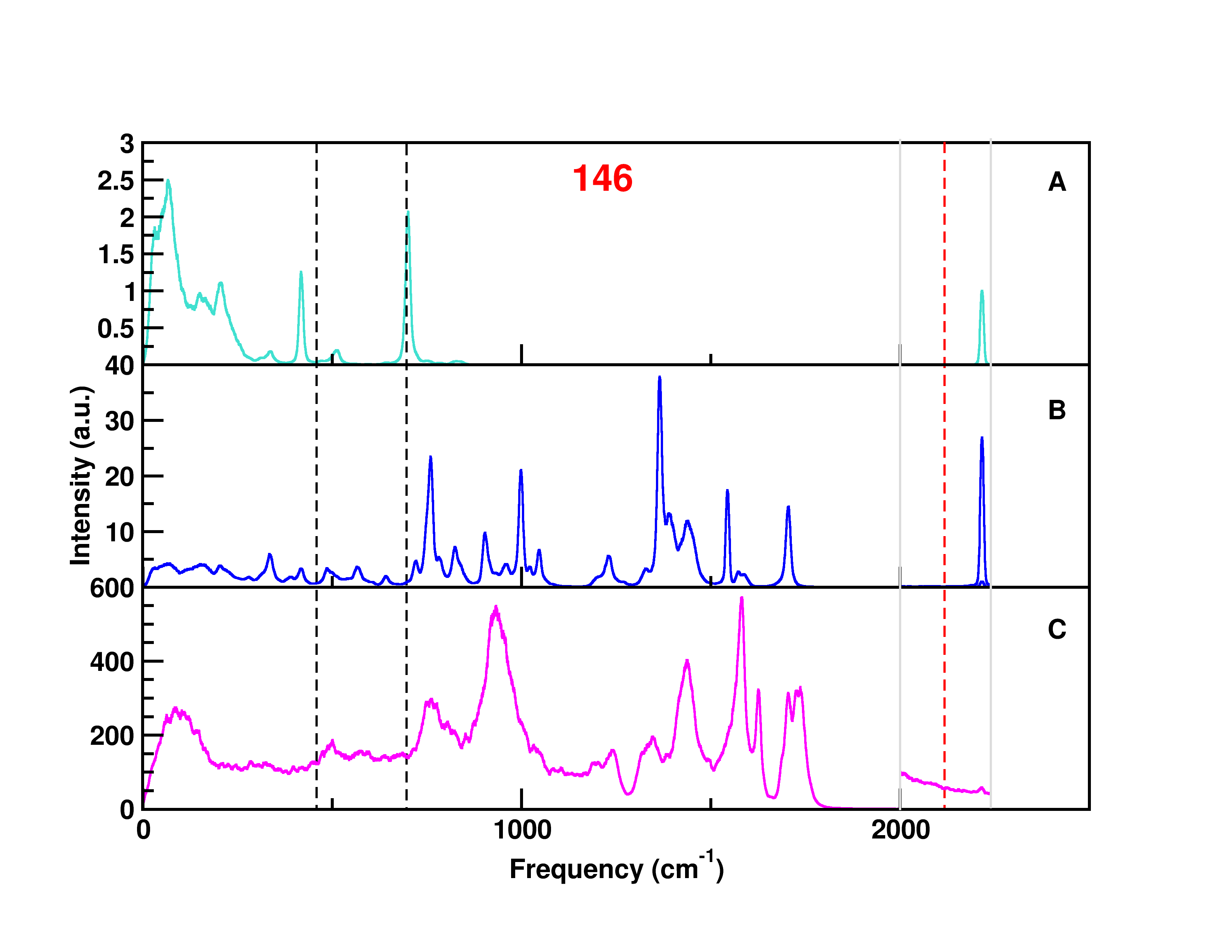}
  \label{fig:ala146}
\end{minipage}
\end{figure}
\begin{figure}
\begin{minipage}{.5\textwidth}
  \centering
  \includegraphics[width=1.1\linewidth]{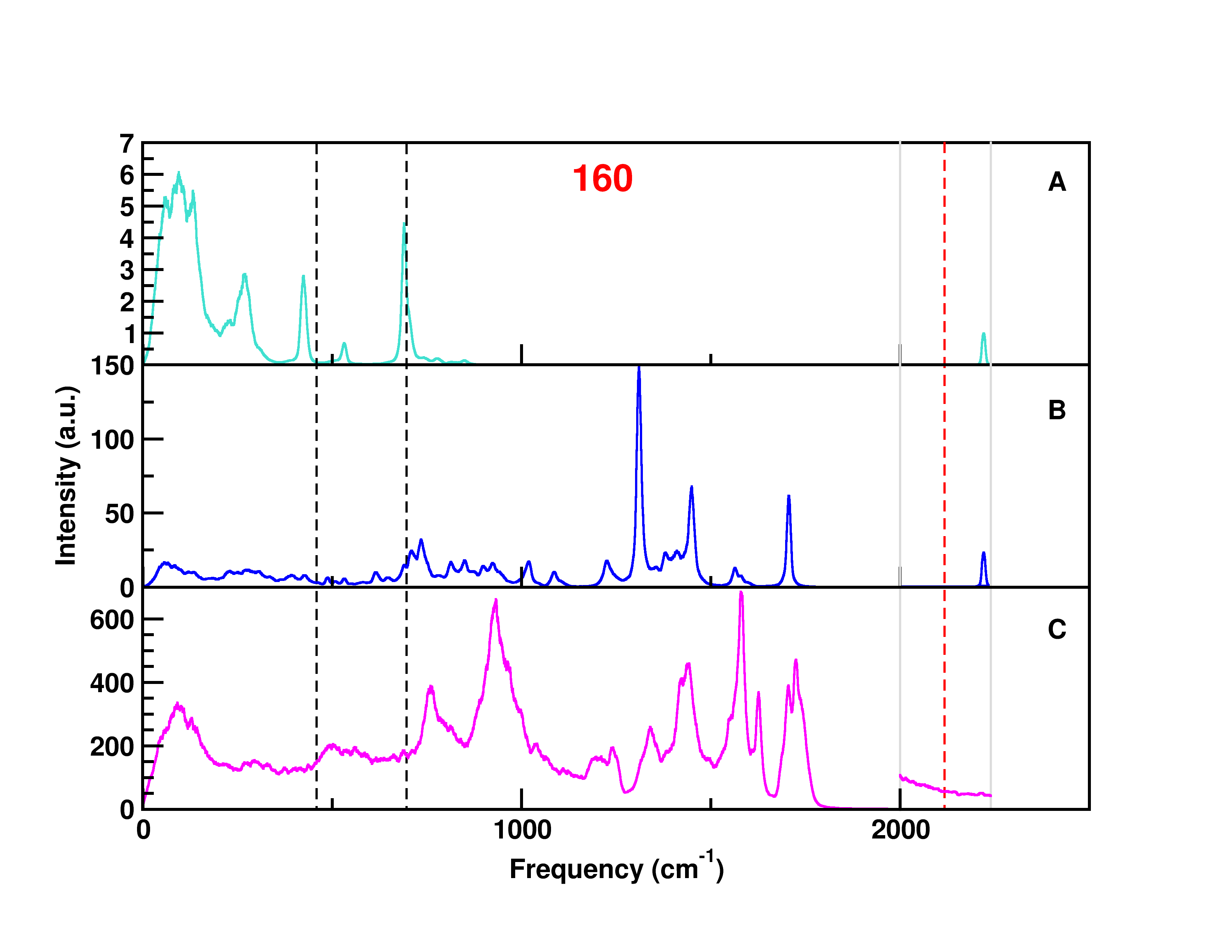}
  \label{fig:ala160}
\end{minipage}
\caption{IR spectra of SCN label (a), labelled alanine residue (b),
  and full lysozyme (c) for all 16 SCN-attached alanine residues in
  lysozyme. The dashed lines denote the specific peaks of SCN label
  MeSCN at 459, 697, and 2118 cm$^{-1}$. The range highlighted with
  grey lines between 2000 and 2240 cm$^{-1}$ is rescaled.}
\label{sifig:ala_3ir}
\end{figure}

\clearpage

\begin{figure}[H]
\begin{center}
\includegraphics[width=0.8\textwidth]{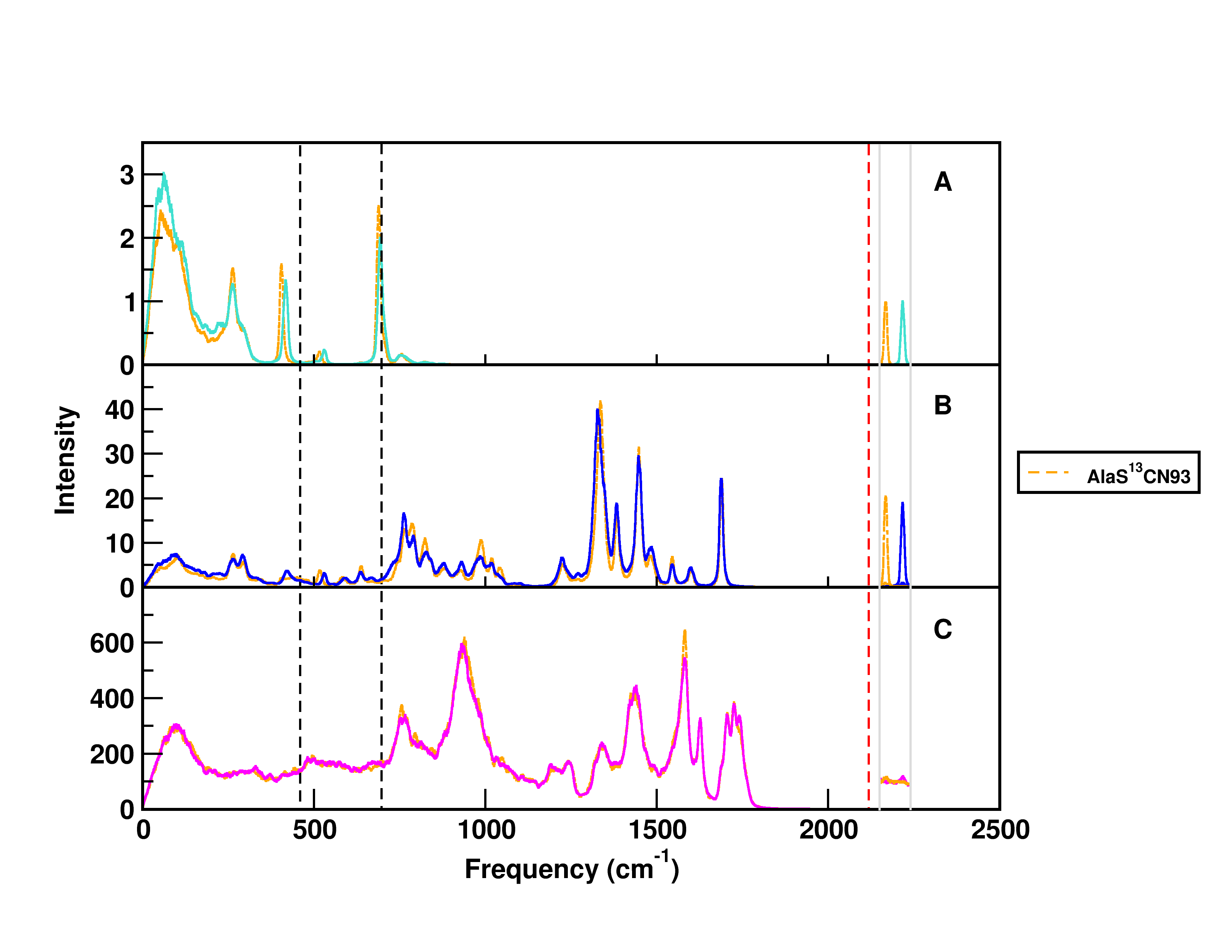}
\caption{IR spectra of SCN label and isotopically S$^{13}$CN label
  (a), labelled alanine residue and isotopically S$^{13}$CN labelled
  alanine residue (b), and full lysozyme and isotopically S$^{13}$CN
  labelled lysozyme (c) for Ala93SCN. The orange dashed line in each
  panel represents IR spectra containing isotopically S$^{13}$CN
  label. The dashed lines denote the specific peaks of SCN label MeSCN
  at 459, 697, and 2118 cm$^{-1}$. The range illustrated with grey
  lines between 2150 and 2240 cm$^{-1}$ is rescaled.}
\label{sifig:iso_alascn93}
\end{center}
\end{figure}

\begin{figure}[H]
\begin{center}
\includegraphics[width=0.8\textwidth]{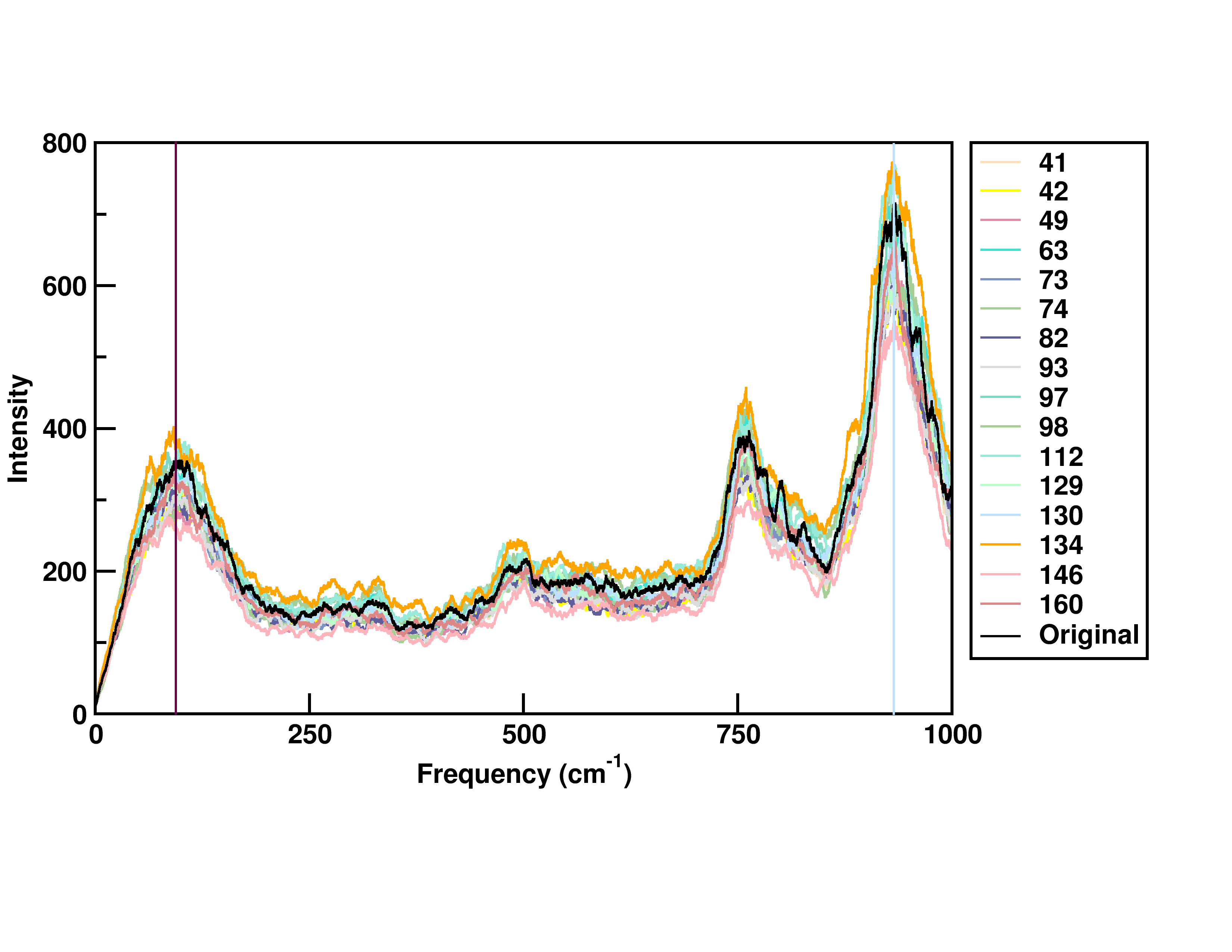}
\caption{IR spectra of lysozyme for all 16 AlaQQSCN residues with
  original lysozyme at low frequency.}
\label{sifig:ir_org}
\end{center}
\end{figure}

\clearpage

\bibliography{refs}

\end{document}